\documentclass[journal,10pt]{IEEEtran}

\usepackage{cite}
\usepackage{graphicx}
\usepackage{amsmath}
\usepackage{array}
\usepackage{tikz}
\usepackage{mdwmath}
\usepackage{amssymb}
\usepackage{mdwtab}
\usepackage{stfloats}
\usepackage[tight,footnotesize]{subfigure}
\usepackage{amsmath,amsthm}
\usepackage{threeparttable}
\usepackage{color}
\usepackage{url}
\usepackage{algpseudocode}
\usepackage{algorithm}
\usepackage{multicol}
\usepackage{amssymb}
\usepackage{epstopdf}

\algrenewcommand{\algorithmicrequire}{\textbf{Input:}}
\algrenewcommand{\algorithmicensure}{\textbf{Output:}}

\newtheorem{theorem}{Theorem}

\newtheorem{lemma}{Lemma}

\usetikzlibrary{shapes.geometric, arrows}
\begin{document}
\title{On the Performance of RIS-Aided Spatial Modulation for Downlink Transmission}
\author{
Xusheng Zhu,  Qingqing Wu, \IEEEmembership{Senior Member, IEEE}, and Wen Chen, \IEEEmembership{Senior Member, IEEE}

\thanks{
Part of this work was accepted by IEEE ICC workshop 2024 \cite{zhupere2024}.
The work of Qingqing Wu was supported by National Key R\&D Program of China (2023YFB2905000), NSFC 62371289, and NSFC 62331022.
The work of Wen Chen was supported by National key project 2020YFB1807700, NSFC 62071296, Shanghai Kewei 22JC1404000.
(\emph{Corresponding author: Qingqing Wu}.)
}
\thanks{Xusheng Zhu, Qingqing Wu, and Wen Chen are with the Department of Electronic Engineering, Shanghai Jiao Tong University, Shanghai 200240, China (e-mail: xushengzhu@sjtu.edu.cn; qingqingwu@sjtu.edu.cn; wenchen@sjtu.edu.cn).}
}

\maketitle
\begin{abstract}
In this study, we explore the performance of a reconfigurable reflecting surface (RIS)-assisted transmit spatial modulation (SM) system for downlink transmission, wherein the deployment of RIS serves the purpose of blind area coverage within the channel. At the receiving end, we present three detectors, i.e.,  maximum likelihood (ML) detector, two-stage ML detection, and greedy detector to recover the transmitted signal. By utilizing the ML detector, we initially derive the conditional pair error probability expression for the proposed scheme. Subsequently, we leverage the central limit theorem (CLT) to obtain the probability density function of the combined channel. Following this, the Gaussian-Chebyshev quadrature method is applied to derive a closed-form expression for the unconditional pair error probability and establish the union tight upper bound for the average bit error probability (ABEP). Furthermore, we derive a closed-form expression for the ergodic capacity of the proposed RIS-SM scheme. Monte Carlo simulations are conducted not only to assess the complexity and reliability of the three detection algorithms but also to validate the results obtained through theoretical derivation results.

\end{abstract}

\section {Introduction}
With the popularization and commercialization of fifth-generation (5G) mobile technology, the development of sixth-generation (6G) mobile communication technology has been accelerated globally \cite{dang2020what}.
This wireless communications standard is anticipated to support many new services and applications that demand a broad range of enabling technologies\cite{saad2020a}. However, in the face of these new challenges, a new paradigm is urgently required to meet the communication needs of next-generation networks, especially in the physical layer.

One option is reconfigurable intelligent surfaces (RIS), which bring a new degree of regulatory freedom to wireless communication systems by regulating RIS deployed in the channel to control the propagation channel rather than focusing on the transmitter or receiver sides \cite{huang2019recon}.
In particular, RIS is a two-dimensional planar array composed of a large number of low-cost, passive reflective units, which can be adjusted in real-time in terms of phase and/or amplitude shift of the reflective units \cite{chen2023irs}.
Moreover, RIS can provide a power gain of the square of the number of units, which reveals the limit of the performance gain \cite{wu2019inte}.
Given this, the wireless environment can be customized by adjusting the amplitude or phase of the RIS to constructively enhance the desired signal or destructively diminish the interfering signal without any radio-frequency (RF) chains\cite{bash2021reconf,zhang2023inte}.
The aforementioned advantages of RIS have sparked extensive research interest in RIS-assisted wireless communication systems, such as mobile edge computing\cite{chenw2023irs}, Internet-of-Things (IoT)\cite{chen2023act}, and millimeter wave (mmWave) transmission \cite{on2023zhu}.

Another option is spatial modulation (SM), which is a promising modulation technique based on multiple-input multiple-output (MIMO) systems.
To be specific, SM can effectively avoid inter-antenna interference and synchronization problems by activating only one antenna in each transmission time slot\cite{li2021single,mes2008spatial,wen2019a}.
The index of antennas in SM is employed to carry additional information bits through a simple design without additional spectral and energy efficiency \cite{zhu2022on,li2019spati,wen2014per}.
It is worth noting that SM migrates the input information bits to transmit or/and receive antenna indexes in the spatial domain and phase shift keying/quadrature amplitude modulation (PSK/QAM) symbols in the symbol domain, respectively \cite{jeg2003spatial}.
Compared to conventional PSK/QAM modulation, SM can leverage the spatial dimension to enhance the degree of freedom of information modulation\cite{mesleh2015qua}.
It is worth mentioning that antenna selection techniques based on channel state information (CSI) can also alleviate the hardware complexity in MIMO systems\cite{ouyang2024joint}.
To clarify the characteristic of SM and antenna selection schemes, a fair comparison between the two schemes was studied in \cite{gae2019a}.
In particular, \cite{jegan2009space} proposed the space shift keying (SSK) scheme based on SM, in which only the antenna index is applied to transmit the information rather than the transmitted symbols.
Nevertheless, in mmWave frequency bands due to severe path loss, it is difficult for SM to ensure reliable signal transmission with only one antenna. To address this problem, \cite{ding2017spatial} proposed the SSM modulation technique, which achieves reliable transmission by emitting a narrow beam pointing to a specific scatterer by employing the hybrid beam technique \cite{zhu2023qua}.

Due to the above advantages of RIS and SM, there has recently been a great deal of interest in combining RIS with SM and its variants \cite{basar2020rec,ma2020large,luo2021spatial,wu2021rec,zhu2023rissk,bou2021per,zhu2023per,zhu2023RIS,zhudouble23,zhu2024transs,zhu2022erm}.
Specifically,
the authors in \cite{basar2020rec} proposed RIS-SSK and RIS-SM schemes, where RISs are deployed close to the transmitter and spatial domain modulation indices are implemented at the receive antenna.
It was shown that both schemes can not only enhance the average bit error probability (ABEP) performance at extremely low signal-to-noise ratio (SNR) regions but also increase the spectral efficiency of the receive index modulation.
To simultaneously exploit the transmit and receive antenna indices, \cite{ma2020large} investigated the ABEP performance of RIS-SM and derived its analytical expression with the help of an approximating Q-function. In \cite{luo2021spatial}, the RIS-assisted SM uplink communication system is investigated and an optimization problem is formulated to reduce the symbol error rate. It is worth mentioning that \cite{luo2021spatial} studied the two schemes of RIS-assisted transmit SM and receive SM, respectively. Besides, the RIS-assisted SM symbiotic radio scheme was proposed, where the receiver simultaneously detects the data transmitted in the RIS-assisted communication as well as the additional IoT data transmitted by the RIS\cite{wu2021rec}.
In \cite{li2024toward}, the authors investigated a new communication architecture in which a transmission RIS and a transmitter are integrated into the design, featuring a simple structure with low energy consumption.
Considering this,
\cite{zhu2024transs} studied the transmissive RIS assisted SM scheme and derived the closed-form ABEP expression.
Additionally, \cite{zhu2023rissk} investigated the ABEP performance on RIS assisted SSK scheme, where RIS is deployed in the middle of the channel.
Afterward, \cite{zhu2023per} considered a more realistic situation where the channel suffers from estimation errors and investigated its reliability. Meanwhile, \cite{bou2021per} investigated the ABEP performance of the blind RIS-SSK scheme under hardware impairment. In the mmWave band, \cite{zhu2023RIS} proposed the RIS-SSM scheme, where the link from the transmitter to the RIS is transmitted with a line-of-sight (LoS) path, and the link from the RIS to the receiver is conveyed with none-LoS (NLoS) paths. To increase the spectral efficiency of the RIS-SSM scheme, \cite{zhu2022erm} and \cite{zhudouble23} presented the RIS-assisted double SSM scheme that studied the ABEP performance, where transmitter-RIS and RIS-receiver links both apply the NLoS paths to transmit the information.

Against the above background, we propose the RIS-assisted SM downlink multiple-input-single-output (MISO) system, which is a more common transmission mode. To the best of our knowledge, there is no work in the open literature on the RIS-SM scheme for downlink communication systems. In comparison to  \cite{basar2020rec}, we consider a RIS-SM for the downlink model, in which the RIS is placed in the middle of the channel and the SM is implemented at the transmitter, rather than SM is implemented at the receiver side. Unlike \cite{ma2020large}, we study the two metrics ABEP and ergodic capacity separately and derive closed-form expressions accordingly. Different from \cite{luo2021spatial}, we investigate the RIS-SM scheme for the downlink transmission scheme. It is worth mentioning that the signal model considered by \cite{zhu2023rissk,bou2021per,zhu2023per} does not take into account PSK/QAM in the symbol domain, which is a special case of the proposed scheme. The main contributions of this paper are summarized as follows.
\begin{itemize}
\item  In this paper, we investigate a RIS-assisted SM downlink transmission scheme in which a synergistic relationship between SM and RIS-assisted systems is leveraged to improve the blind area coverage of the wireless system. For the proposed scheme, the information bits are mapped not only to the $M$-ary PSK/QAM constellation diagram but also to the corresponding position indices of the transmit antennas.
\item  For the proposed RIS-SM scheme, we employ three detectors to recover the original information, i.e., maximum likelihood (ML) detector, two-stage ML (TSML), and greedy detector (GD). For the ML detector, we perform decoding detection by an exhaustive search of the two-dimensional signal space. For the TSML detector, we first decode the spatial domain signals via the ML detector. Based on this, we decode the symbol domain information by using ML decoding. For the GD, we pick the transmit antenna index with the strongest signal energy to decode the spatial domain information, while for the symbol domain information, we use ML decoding.
\item  Based on the ML detector, we first derive the conditional pair error probability (CPEP) expression for the proposed scheme. Subsequently, we perform the analysis in terms of the two cases of antenna-indexed correct decoding and incorrect decoding, respectively. In this context, we first derive the probability density function (PDF) of the combined channel using the central limit theorem (CLT) and further derive the exact integral expression of the unconditional pair error probability (UPEP) of the proposed scheme based on the CPEP expression. Afterward, the Gaussian-Chebyshev quadrature (GCQ) method is adopted to obtain its closed-form expression. Finally, the joint tight upper bound closed-form expression of ABEP is provided. To gain more insights, we also analyze the ergodic capacity and derive the closed-form expression.
\item  Simulations are provided first to evaluate the computational complexity and reliability of the three detectors. Then, the simulation results are used to verify the correctness of the theoretical derivation.
    The results show that when the number of units in the RIS is not less than 80, the analytical results and simulation results obtained using CLT can match perfectly. Additionally, the impact of errors in the reflection phase shift capability of RIS on ABEP performance was also investigated. Finally, the effects of the number of reflecting units, transmit antenna, and modulation order with respect to ABEP and ergodic capacity performance are also studied.
\end{itemize}

The rest of the paper is organized as follows: Section II describes the considered system model and the three demodulation decoding algorithms. Section III presents the ABEP expression for the RIS-SM scheme and gives the closed-form expression of ergodic capacity. Section IV gives a quantitative analysis of the decoding complexity of the three detectors. Also, the results of ABEP performance and ergodic capacity are provided. Finally, Section V summarizes the whole paper.

\emph {Notations:}
Lowercase bold letters and uppercase bold letters denote vectors and matrices, respectively.
$(\cdot)^T$, $(\cdot)^H$, and $(\cdot)^*$ are the transposition, Hermitian transposition, and complex conjugation of a vector/matrix operations, respectively.
$\mathbb{C}^{L\times M}$ represents the space of $L\times M$ matric.
${\rm diag}(\cdot)$ denotes the diagonal matrix operation.
$\mathcal{N}(\cdot,\cdot)$ and $\mathcal{CN}(\cdot,\cdot)$ denote the real and complex Gaussian distributions, respectively.
$\Pr(\cdot)$, $P_b$, and $\bar P_b$ stand for the probability of the event occurring, CPEP, and UPEP, respectively.
$\Re$ and $\Im$ represent the real and imaginary part operations, respectively.
$E[\cdot]$ and $Var[\cdot]$ indicate the expectation and variance operations, respectively.
The Q-function and Gaussian error function are denoted as $Q(x)=\int_x^\infty\frac{1}{\sqrt{2\pi}}e^{-\frac{1}{2}t^2}dt$ and $\Phi(x)=\frac{2}{\sqrt{\pi}}\int_0^x\exp(-v^2)dv$, respectively.
${\rm MGF}_\Gamma (\cdot)$ represents the moment generating function (MGF) with respect to variable $\Gamma$.

\section {System Model}
\begin{figure}
  \centering
  \includegraphics[width=9cm]{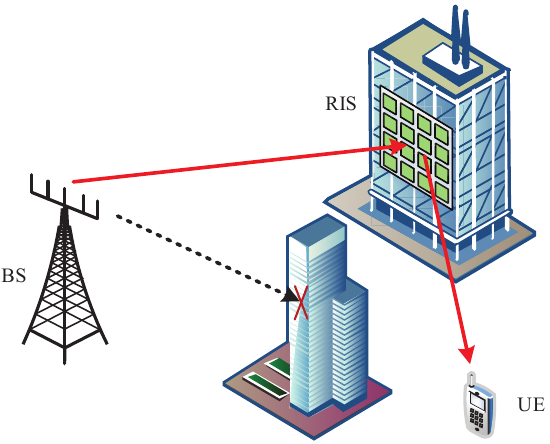}
  \caption{\small{The RIS-assisted SM system model.}}\label{sys}
\end{figure}

In this paper, we consider the RIS-assisted SM downlink transmission system as shown in Fig. \ref{sys}, which is composed of a base station (BS), a user equipment (UE), and a piece of RIS.
Due to the existing building blockage between the BS and UE, we assume that there is no direct link from the BS to UE.
For this reason, we resort to a RIS to assist the downlink signal transmission from the BS to the UE.
In Fig. \ref{sys}, we assume that the BS and UE are equipped with $N_t$ and a single antenna for reception, respectively.
Besides, the RIS consists of $L$ low-cost passive reflection units, each of which can adjust the amplitude and phase of the reflected signal independently.
To characterize the performance limit of RIS, we assume that the reflection amplitude per unit number is one.
In the following, we utilize the matrix $\boldsymbol{\Theta} ={\rm diag}\{e^{j\theta_1},e^{j\theta_2},\cdots,e^{j\theta_L}\}$ to represent RIS reflection coefficient\cite{zhu20203fd,zhu2023sic,zhu2024robust}, where $\theta_l, l\in\{1,2\cdots,L\}$ is the reflection phase shift of the $l$-th unit on the RIS.

The wireless channels between the UE and RIS, and RIS and $n_t$-th antenna of BS are respectively described as $\boldsymbol{g}\in\mathbb{C}^{L\times 1}$ and
$\boldsymbol{h}_{n_t}\in\mathbb{C}^{L\times 1}$, where $g_l =\beta_le^{-j\phi_l}$ and $h_{l,n_t} =\alpha_{l,n_t}e^{-j\varphi_{l,n_t}}$ denote the channels between the $l$-th reflecting element of RIS and UE and $l$-th reflecting element and BS, respectively.
Note that $\beta_l$ and $\phi_l$, $\alpha_{l,n_t}$ and ${\varphi_{l,n_t}}$ represent the amplitudes and phases of the fading channels $g_l$ and $h_{l,n_t}$, respectively.
To explore the achievable performance of SM, we assume that the BS and UE can obtain perfect CSI. For the required reflection coefficient, the BS sends it to the RIS controller through a wired or wireless link. It is worth noting that the RIS controller can tune the reflection state of each RIS element in real time based on the information received from the BS side.

\subsection{Signal Model of RIS-SM Scheme}
At each time slot, the input message $\log_2(N_t)+\log_2(M)$ consists of two parts. The first $\log_2(M)$ bit maps to a constellation point $s$ in the $M$-PSK/QAM signal selected from the symbol domain code book $\mathcal{S}= \{s_1,s_2,\cdots,s_M\}$, where the corresponding average power with respect to $s$ satisfies $E[|s|^2]=1$.
The rest of the $\log_2(N_t)$ bits come from the spatial domain constellation point, which is selected by the spatial domain index set $\mathcal{N}=\{1,2,\cdots,{N_t}\}$.
In this way, the received signal at the UE side is given by
\begin{equation}\label{received}
\begin{aligned}
y &= \sqrt{P_t}\boldsymbol{g\Theta}\boldsymbol{h}_{n_t}s + n_0\\
&=\sqrt{P_t}\sum\nolimits_{l=1}^L\alpha_{l,n_t}\beta_le^{j(\theta_l-\phi_l-\varphi_{l,n_t})}s + n_0,
\end{aligned}
\end{equation}
where $P_t$ denotes the average transmit power and $n_0\sim\mathcal{CN}(0,N_0)$ represents the additive white Gaussian noise (AWGN).
In this manner, the instantaneous SNR of RIS-SM scheme can be defined as
\begin{equation}
{\rm SNR} = \frac{{P_t}\left|\sum_{n=1}^N\alpha_{n_t,l}\beta_le^{j(\theta_l-\phi_l-\varphi_{l,n_t})}\right|^2}{N_0}.
\end{equation}
In the case of known perfect CSI, the RIS can be adjusted by adjusting the reflected phase shift, thus canceling out the phase information of the received and reflected channels.
As such, we have $\theta_l=\phi_l+\varphi_{l,n_t}$.
Here, (\ref{received}) can be further given by
\begin{equation}
\begin{aligned}
y =\sqrt{P_t}\sum\nolimits_{l=1}^L\alpha_{l,n_t}\beta_ls + n_0.
\end{aligned}
\end{equation}

\subsection {Signal Detection of RIS-SM Scheme}
In this subsection, ML, TSML, and GD detection algorithms are used to estimate the spatial and symbol domain information in RIS-SM, respectively.

\subsubsection{ML Detector}
At the UE of RIS-SM system, the ML detector jointly estimates the transmit antenna index $\hat n_t$ and the symbol information $s$. It is worth noting that the ML detector exhaustively searches all combinations of $n_t$ and $s$ and differs all combination results from the emission information, where the $n_t$ and $s$ value with the smallest result is designated as the detection value. Based on this, the ML detector of the RIS-SM system is defined as follows
\begin{equation}\label{ml}
[\hat n_t, \hat s] = \arg\min\limits_{n_t\in\mathcal{N}, s\in\mathcal{S}}\left|y-\sqrt{P_t}\sum\nolimits_{l=1}^{L}\alpha_{l,n_t}\beta_l s\right|^2.
\end{equation}
\subsubsection{TSML Detector}
For the above ML detection algorithm, the approach of employing exhaustive traversal leads to high complexity. In order to reduce its detection complexity, we adopt a two-stage detection approach. Firstly, ML detection is performed on the signal in the spatial domain, which is expressed as follows
\begin{equation}\label{tsml}
[\hat n_t] = \arg\min\limits_{n_t\in\mathcal{N}}\left|y-\sqrt{P_t}\sum\nolimits_{l=1}^{L}\alpha_{l,n_t}\beta_l \tilde s\right|^2,
\end{equation}
where $\tilde s$  denotes an arbitrary signal in $\mathcal{S}$ space.
Using the obtained $\hat n_t$, we then check all possibilities through minimum Euclidean distance to estimate symbol domain information as follows:
\begin{equation}\label{sicsymbol}
[\hat s] = \arg\min\limits_{s\in\mathcal{S}}\left|y-\sqrt{P_t}\sum\nolimits_{l=1}^{L}\alpha_{l,\hat n_t}\beta_l s\right|^2.
\end{equation}

\subsubsection{GD}
According to the phase shift adjustment mechanism of RIS, the phase of the transmitted signal from the transmit antenna and the reflected signal arriving at UE is eliminated, thereby maximizing SNR. On the contrary, the SNR of the remaining transmit antennas reflected from RIS to UE is not the largest. Therefore, we can use the GD algorithm to select the antenna index with the highest instantaneous energy as the estimated antenna index, where the antenna index is estimated based on the received signal as
\begin{equation}\label{gd1}
\hat n_t =\arg\max\limits_{n_t\in\mathcal{N}}|y|^2.
\end{equation}
Taking advantage of the detected antenna index obtained from (\ref{gd1}), we estimate symbol signal similarly as (\ref{sicsymbol}).

\section {Performance Analysis}
In this section, the analytical expressions of ABEP are derived over the Rayleigh fading channels based on the ML detector.
\subsection{CPEP}
To gain an upper bound on ABEP, we first derive the CPEP expression as
\begin{equation}\label{cpep1}
\begin{aligned}
P_b 
=&\Pr\left(|y-\sqrt{P_t}\sum\limits_{l=1}^{L}\alpha_{l,n_t}\beta_l s|^2\right.\\&\left.>|y-\sqrt{P_t}\sum\limits_{l=1}^L\alpha_{l,\hat n_t}\beta_{l}e^{-j(\theta_{l,\hat n_t}-\theta_{l,n_t})}\hat s|^2\right)\\
\overset{(a)}{=}&\Pr(|y-\sqrt{P_t}G_{n_t} s|^2>|y-\sqrt{P_t}G_{\hat n_t}\hat s|^2)\\
=&\Pr(\left|n_0\right|^2>|\sqrt{P_t}(G_{ n_t} s-G_{\hat n_t}\hat s)+n_0|^2)\\
=&\Pr(-|\sqrt{P_t}\left(G_{ n_t} s-G_{\hat n_t}\hat s\right)|^2\\&-2\Re\{n_0^H\sqrt{P_t}\left(G_{ n_t} s-G_{\hat n_t}\hat s\right)\}>0)\\
=&\Pr\left(F>0\right),
\end{aligned}
\end{equation}
where $(a)$ denotes the $G_{n_t}=\sum_{l=1}^{L}\alpha_{l,n_t}\beta_l$ and $G_{\hat n_t}=\sum_{l=1}^L\alpha_{l,\hat n_t}\beta_{l}e^{-j(\theta_{l,\hat n_t}-\theta_{l,n_t})}$.
Note that $F=-\left|\sqrt{P_t}\left(G_{ n_t} s-G_{\hat n_t}\hat s\right)\right|^2-2\Re\{n_0^H\sqrt{P_t}\left(G_{ n_t} s-G_{\hat n_t}\hat s\right)\}$ is a real Gaussian random variable\footnote{In each transmission time slot, the transmit signal $s$ is a deterministic signal. As the signal propagates through the channel to reach the receiver, AWGN $n_0$ is introduced. In the case of known perfect CSI, the noise signal is a random variable satisfying a complex Gaussian distribution, and the remaining quantities in (\ref{cpep1}) are known constant values.
Based on this, the $F$ in (\ref{cpep1}) follows the real Gaussian random variable.}.
Accordingly, the mean and variance values of $F$ can be given as
$
\mu_F=-\left|\sqrt{P_t}\left(G_{ n_t} s-G_{\hat n_t}\hat s\right)\right|^2$ and $
\sigma_F^2=2N_0P_t\left|G_{ n_t} s-G_{\hat n_t}\hat s\right|^2
$
Herein, the CPEP can be calculated as
\begin{equation}\label{cpep2}
P_b=Q(-\mu_F/\sigma_F^2)=Q\left(\sqrt{\frac{\rho\left|G_{ n_t} s-G_{\hat n_t}\hat s\right|^2}{2}}\right),
\end{equation}
where $\rho = {P_t}/{N_0}$ stands for the average SNR.
\subsection{UPEP}
To obtain the unconditional PEP, we then consider two cases of correct transmit antenna detection, i.e., $\hat n_t = n_t$, and erroneous transmit antenna detection, i.e., $\hat n_t \neq n_t$, as follows:
\subsubsection{Correct transmit antenna detection $\hat n_t = n_t$}
In this case, the (\ref{cpep2}) can be simplified as
\begin{equation}
P_b=Q\left(\sqrt{\frac{\rho|G_{ n_t}|^2| s-\hat s|^2}{2}}\right).
\end{equation}
For the sake of simplicity, we define $x=|G_{ n_t}|^2$.
In this way, the UPEP can be given as
\begin{equation}\label{cpep3}
\begin{aligned}
\bar P_b &= \int_0^\infty f(x)Q\left(\sqrt{\frac{\rho x|s-\hat s|^2}{2}}\right)dx,
\end{aligned}
\end{equation}
where $f(x)$ denotes the PDF with respect to the variable $x$. Since it is difficult to find the distribution of $x$ directly, we first calculate the distribution of $\xi = \sum_{l=1}^{L}\alpha_{l,n_t}\beta_l$. Note that the distribution of $f(x)$ is obtained on the basis of the distribution of $\xi$ provided in {\bf Theorem 1}.
\begin{theorem}
The distribution of variable $\xi$ can be described as
\begin{equation}
\xi\sim\mathcal{N}\left(\mu_\xi, \sigma_\xi^2\right),
\end{equation}
where the mean and variance can be denoted as $\mu_\xi=\frac{L\pi}{4}$ and $\sigma_\xi^2=\frac{L(16-\pi^2)}{16}$, respectively.
\end{theorem}

{\emph {Proof:}} Please refer to Appendix A.
$\hfill\blacksquare$

From the distribution given by {\bf Theorem 1}, one can further derive the PDF of $x$ in {\bf Theorem 2}.
\begin{theorem}
The PDF of variable $x$ can be written as
\begin{equation}\label{xpdf}
f(x) = \frac{\exp\left(-\frac{x+\mu_\xi^2}{2\sigma_\xi^2}\right)}{2\sqrt{2\pi\sigma_\xi^2x}}\left[\exp\left(\frac{\mu_\xi\sqrt{x}}{\sigma_\xi^2}\right)+\exp\left(-\frac{\mu_\xi\sqrt{x}}{\sigma_\xi^2}\right)\right].
\end{equation}
\end{theorem}

{\emph {Proof:}} Please refer to Appendix B.
$\hfill\blacksquare$

Substituting (\ref{xpdf}) into (\ref{cpep3}), we can rewrite (\ref{cpep3}) as

\begin{equation}\label{xpdf1}
\begin{aligned}
\bar P_b
=&\frac{1}{2\sqrt{2\pi\sigma_\xi^2}}\int_0^\infty \frac{\exp\left(-\frac{x+\mu_\xi^2}{2\sigma_\xi^2}\right)}{\sqrt{x}}\left[\exp\left(\frac{\mu_\xi\sqrt{x}}{\sigma_\xi^2}\right)\right.\\&\left.+\exp\left(-\frac{\mu_\xi\sqrt{x}}{\sigma_\xi^2}\right)\right]Q\left(\sqrt{\frac{\rho x|s-\hat s|^2}{2}}\right)dx.
\end{aligned}
\end{equation}
By using the equivalent form of $Q(x)=\frac{1}{\pi}\int_0^{\frac{\pi}{2}}\exp\left(-\frac{x^2}{2\sin^2\theta}\right)d\theta$, then (\ref{xpdf1}) can be rephrased as
\begin{equation}
\begin{aligned}
\bar P_b
=&\frac{1}{2\pi\sqrt{2\pi\sigma_\xi^2}}\int_0^\infty \int_0^{\frac{\pi}{2}} \frac{\exp\left(-\frac{x+\mu_\xi^2}{2\sigma_\xi^2}\right)}{\sqrt{x}}\left[\exp\left(\frac{\mu_\xi\sqrt{x}}{\sigma_\xi^2}\right)\right.\\&\left.+\exp\left(-\frac{\mu_\xi\sqrt{x}}{\sigma_\xi^2}\right)\right]\exp\left(-\frac{\rho |s-\hat s|^2x}{4\sin^2\theta}\right)d\theta dx.
\end{aligned}
\end{equation}
Since the above equation is hard to unlock directly, we exchange the order of its integrals to get
\begin{equation}\label{xpdf2}
\begin{aligned}
\bar P_b
=&\frac{1}{2\pi\sqrt{2\pi\sigma_\xi^2}}\int_0^{\frac{\pi}{2}} \int_0^\infty \frac{\exp\left(-\frac{x+\mu_\xi^2}{2\sigma_\xi^2}\right)}{\sqrt{x}}\left[\exp\left(\frac{\mu_\xi\sqrt{x}}{\sigma_\xi^2}\right)\right.\\&\left.+\exp\left(-\frac{\mu_\xi\sqrt{x}}{\sigma_\xi^2}\right)\right]\exp\left(-\frac{\rho |s-\hat s|^2x}{4\sin^2\theta}\right)dxd\theta.
\end{aligned}
\end{equation}
Without loss of generality, we let $t = \sqrt{x}$. In this manner, (\ref{xpdf2}) can be rewritten as.
\begin{equation}\label{xpdf3}
\begin{aligned}
\bar P_b
=&\frac{\exp\left(-\frac{\mu_\xi^2}{2\sigma_\xi^2}\right)}{\pi\sqrt{2\pi\sigma_\xi^2}}\int_0^{\frac{\pi}{2}} \int_0^\infty \exp\left(-\frac{t^2}{2\sigma_\xi^2}\right)\left[\exp\left(\frac{\mu_\xi t}{\sigma_\xi^2}\right)\right.\\&\left.+\exp\left(-\frac{\mu_\xi t}{\sigma_\xi^2}\right)\right]\exp\left(-\frac{\rho |s-\hat s|^2t^2}{4\sin^2\theta}\right)dtd\theta.
\end{aligned}
\end{equation}
After some mathematical manipulations, we can recast (\ref{xpdf3}) as
\begin{equation}\label{xpdf31}
\begin{aligned}
\bar P_b
&=\frac{\exp\left(-\frac{\mu_\xi^2}{2\sigma_\xi^2}\right)}{\pi\sqrt{2\pi\sigma_\xi^2}}\int_0^{\frac{\pi}{2}} \int_0^\infty \exp\left(-\frac{2\sin^2\theta+\rho\sigma_\xi^2 |s-\hat s|^2}{4\sigma_\xi^2\sin^2\theta}t^2\right.\\&\left.+\frac{\mu_\xi t}{\sigma_\xi^2}\right)+\exp\left(-\frac{2\sin^2\theta+\rho\sigma_\xi^2 |s-\hat s|^2}{4\sigma_\xi^2\sin^2\theta}t^2-\frac{\mu_\xi t}{\sigma_\xi^2}\right)dtd\theta.
\end{aligned}
\end{equation}
To avoid redundancy operations, we define $A_1$ and $A_2$ as
\begin{subequations}\label{xpdf4}
\begin{align}
A_1&=\int_0^\infty \exp\left(-\frac{2\sin^2\theta+\rho\sigma_\xi^2 |s-\hat s|^2}{4\sigma_\xi^2\sin^2\theta}t^2+\frac{\mu_\xi t}{\sigma_\xi^2}\right)dt,\\
A_2&=\int_0^\infty \exp\left(-\frac{2\sin^2\theta+\rho\sigma_\xi^2 |s-\hat s|^2}{4\sigma_\xi^2\sin^2\theta}t^2-\frac{\mu_\xi t}{\sigma_\xi^2}\right)dt.
\end{align}
\end{subequations}
Fortunately, we can evaluate (\ref{xpdf4}) directly from the equation provided by \cite{jeff2007}
\begin{equation}\label{xpdf5}
\begin{aligned}
\int_0^\infty\exp\left(-\frac{t^2}{4\delta}-\gamma t\right)dt=\sqrt{\pi \delta}\exp(\delta\gamma^2)[1-\Phi(\gamma\sqrt{\delta})].
\end{aligned}
\end{equation}
With the help of (\ref{xpdf5}), then (\ref{xpdf4}) can be recast as
\begin{subequations}\label{xpdf6}
\begin{align}
\nonumber A_1&=\sqrt{\frac{\pi\sigma_\xi^2\sin^2\theta}{2\sin^2\theta+\rho\sigma_\xi^2 |s-\hat s|^2}}\\
&\times\exp\left(
\frac{\mu_\xi^2\sin^2\theta}{(2\sin^2\theta+\rho\sigma_\xi^2 |s-\hat s|^2)\sigma_\xi^2}
\right)\\&\times\left[1-\Phi\left(-\sqrt{\frac{\mu_\xi^2\sin^2\theta}{(2\sin^2\theta+\rho\sigma_\xi^2 |s-\hat s|^2)\sigma_\xi^2}}\right)\right],\nonumber\\
\nonumber A_2
&=\sqrt{\frac{\pi\sigma_\xi^2\sin^2\theta}{2\sin^2\theta+\rho\sigma_\xi^2 |s-\hat s|^2}}\\
&\times\exp\left(
\frac{\mu_\xi^2\sin^2\theta}{(2\sin^2\theta+\rho\sigma_\xi^2 |s-\hat s|^2)\sigma_\xi^2}
\right)\\
&\times\left[1-\Phi\left(\sqrt{\frac{\mu_\xi^2\sin^2\theta}{(2\sin^2\theta+\rho\sigma_\xi^2 |s-\hat s|^2)\sigma_\xi^2}}\right)\right].\nonumber
\end{align}
\end{subequations}
Considering this, the sum of $A_1$ and $A_2$ can be expressed as
\begin{equation}\label{xpdf7}
\begin{aligned}
&A_1+A_2=\sqrt{\frac{\pi\sigma_\xi^2\sin^2\theta}{2\sin^2\theta+\rho\sigma_\xi^2 |s-\hat s|^2}}\\&\times\exp\left(
\frac{\mu_\xi^2\sin^2\theta}{(2\sin^2\theta+\rho\sigma_\xi^2 |s-\hat s|^2)\sigma_\xi^2}
\right)\\&\times\left[2-\Phi\left(-\sqrt{\frac{\mu_\xi^2\sin^2\theta}{(2\sin^2\theta+\rho\sigma_\xi^2 |s-\hat s|^2)\sigma_\xi^2}}\right)\right.\\&\left.-\Phi\left(\sqrt{\frac{\mu_\xi^2\sin^2\theta}{(2\sin^2\theta+\rho\sigma_\xi^2 |s-\hat s|^2)\sigma_\xi^2}}\right)\right].
\end{aligned}
\end{equation}
\begin{lemma}
For the two terms in (\ref{xpdf7}) with respect to $\Phi(\cdot)$ function, we have
\begin{equation}
\begin{aligned}
&\Phi\left(-\sqrt{\frac{\mu_\xi^2\sin^2\theta}{(2\sin^2\theta+\rho\sigma_\xi^2 |s-\hat s|^2)\sigma_\xi^2}}\right)\\&+\Phi\left(\sqrt{\frac{\mu_\xi^2\sin^2\theta}{(2\sin^2\theta+\rho\sigma_\xi^2 |s-\hat s|^2)\sigma_\xi^2}}\right)=0.
\end{aligned}
\end{equation}
\end{lemma}

\emph{ Proof:} Please refer to Appendix C.
$\hfill\blacksquare$

By utilizing {\bf Lemma 1}, the (\ref{xpdf7}) can be simplified to
\begin{equation}\label{xpdf8}
\begin{aligned}
A_1+A_2=&\sqrt{\frac{4\pi\sigma_\xi^2\sin^2\theta}{2\sin^2\theta+\rho\sigma_\xi^2 |s-\hat s|^2}}\\&\times\exp\left(
\frac{\mu_\xi^2\sin^2\theta}{(2\sin^2\theta+\rho\sigma_\xi^2 |s-\hat s|^2)\sigma_\xi^2}
\right).
\end{aligned}
\end{equation}
By replacing (\ref{xpdf8}) with (\ref{xpdf31}), the UPEP can be re-expressed as
\begin{equation}
\begin{aligned}
\bar P_b
=&\frac{\exp\left(-\frac{\mu_\xi^2}{2\sigma_\xi^2}\right)}{\pi\sqrt{2\pi\sigma_\xi^2}}\int_0^{\frac{\pi}{2}} \sqrt{\frac{4\pi\sigma_\xi^2\sin^2\theta}{2\sin^2\theta+\rho\sigma_\xi^2 |s-\hat s|^2}}\\&\times\exp\left(
\frac{\mu_\xi^2\sin^2\theta}{(2\sin^2\theta+\rho\sigma_\xi^2 |s-\hat s|^2)\sigma_\xi^2}
\right) d\theta.
\end{aligned}
\end{equation}
After some manipulations, we have
\begin{equation}\label{epb1}
\begin{aligned}
\bar P_b
=&\frac{1}{\pi}\int_0^{\frac{\pi}{2}} \sqrt{\frac{2\sin^2\theta}{2\sin^2\theta+\rho\sigma_\xi^2 |s-\hat s|^2}}\\&\times\exp\left(
\frac{-\rho\mu_\xi^2|s-\hat s|^2}{4\sin^2\theta+2\rho\sigma_\xi^2 |s-\hat s|^2}
\right) d\theta.
\end{aligned}
\end{equation}
Without loss of generality, let us define $\theta=\frac{\pi}{4}\varpi+\frac{\pi}{4}$, then the (\ref{epb1}) can be recast as
\begin{equation}\label{epb2}
\begin{aligned}
\bar P_b
=&\frac{1}{4}\int_{-1}^1 \sqrt{\frac{2\sin^2\left(\frac{\pi}{4}\varpi+\frac{\pi}{4}\right)}{2\sin^2\left(\frac{\pi}{4}\varpi+\frac{\pi}{4}\right)+\rho\sigma_\xi^2 |s-\hat s|^2}}\\&\times\exp\left(
\frac{-\rho\mu_\xi^2|s-\hat s|^2}{4\sin^2\left(\frac{\pi}{4}\varpi+\frac{\pi}{4}\right)+2\rho\sigma_\xi^2 |s-\hat s|^2}
\right) d\varpi.
\end{aligned}
\end{equation}
Due to the distribution of integral variables in (\ref{epb2}), it is very challenging to address it directly. To tackle this issue, we utilize the GCQ approach\footnote{A detailed description of the accuracy and assessment error of the GCQ approach to UPEP assessment is provided in \cite{zhu20203fd,zhu2023sic,zhu2024robust}.}. Let us define
$\varpi=\cos\left(\frac{2q-1}{2Q}\pi\right)$, $\bar P_b$ can be further evaluated as
\begin{equation}
\begin{aligned}
&\bar P_b
=\frac{\pi}{4Q}\sum_{q=1}^Q \sqrt{\frac{2\sin^2\left(\frac{\pi}{4}\cos\left(\frac{2q-1}{2Q}\pi\right)+\frac{\pi}{4}\right)}{2\sin^2\left(\frac{\pi}{4}\cos\left(\frac{2q-1}{2Q}\pi\right)\!+\!\frac{\pi}{4}\right)\!+\!\rho\sigma_\xi^2 |s-\hat s|^2}}\\&\times\exp\!\left(
\frac{-\rho\mu_\xi^2|s-\hat s|^2}{4\sin^2\left(\frac{\pi}{4}\cos\left(\frac{2q-1}{2Q}\pi\right)\!+\!\frac{\pi}{4}\right)\!+\!2\rho\sigma_\xi^2 |s\!-\!\hat s|^2}
\right) \!+\!R_Q,
\end{aligned}
\end{equation}
where $Q$ is the complexity-accuracy tradeoff factor, and $R_Q$ denotes the error term which can be neglected as the value of $Q$ is large.
\subsubsection{ Erroneous transmit antenna detection  $\hat n_t \neq n_t$}

Based on (\ref{cpep2}), the UPEP can be written as
\begin{equation}\label{cpep3n1}
\begin{aligned}
\bar P_b &= \int_0^\infty f(x)Q\left(\sqrt{\frac{\rho\left|G_{ n_t} s-G_{\hat n_t}\hat s\right|^2}{2}}\right)dx\\
&= \frac{1}{\pi}\int_0^\infty \int_0^{\frac{\pi}{2}} f(x)\exp\left(-\frac{\rho \left|G_{ n_t} s-G_{\hat n_t}\hat s\right|^2}{4\sin^2\theta}\right)d\theta dx\\
&= \frac{1}{\pi}\int_0^{\frac{\pi}{2}}\int_0^\infty f(x)\exp\left(-\frac{\rho \left|G_{ n_t} s-G_{\hat n_t}\hat s\right|^2}{4\sin^2\theta}\right)dxd\theta\\
&= \frac{1}{\pi}\int_0^{\frac{\pi}{2}}{\rm MGF}_\Gamma \left(\frac{-\rho }{4\sin^2\theta}\right) d\theta,
\end{aligned}
\end{equation}
where $f(x)$ denotes the PDF with respect the channel part in $\left|G_{ n_t} s-G_{\hat n_t}\hat s\right|^2$.

Although the integral variable in (\ref{cpep3n1}) contains only the channel component, the information in SM consists of two parts, the spatial domain and the symbol domain. Unfortunately, these two parts are coupled with each other in (\ref{cpep3n1}), which is challenging to resolve.
Nevertheless, in the following, we separate the information into two parts, real and imaginary, to address (\ref{cpep3n1}). Before that, we define
\begin{equation}\label{cpep3n2}
\begin{aligned}
\Gamma &= \left|G_{n_t}s-G_{\hat n_t}\hat s\right|^2\\
&=\left|\sum\limits_{l=1}^{L}\alpha_{l,n_t}\beta_ls-\sum\limits_{l=1}^L\alpha_{l,\hat n_t}\beta_{l}e^{-j(\theta_{l,\hat n_t}-\theta_{l,n_t})}\hat s\right|^2\\
&=\left|\sum\limits_{l=1}^{L}\beta_l\left(\alpha_{l,n_t}s-\alpha_{l,\hat n_t}e^{-j\phi_l}\hat s\right)\right|^2,
\end{aligned}
\end{equation}
where $\phi_l = \theta_{l,\hat n_t}-\theta_{l,n_t}$.
It can be observed that $s$, $e^{-j\phi_l}$, and $\hat s$ in (\ref{cpep3n2}) are complex values.
In this respect, we express (\ref{cpep3n2}) in terms of real and imaginary parts as
\begin{equation}
\begin{aligned}
&\Gamma
=\left|\sum\limits_{l=1}^{L}\beta_l
\left(\alpha_{l,n_t}s_\Re-\alpha_{l,\hat n_t}(\cos\phi_l\hat s_\Re+\sin\phi_l\hat s_\Im)\right)
\right.
\\&
\left.
+j\sum\limits_{l=1}^{L}\beta_l\left(\alpha_{l,n_t}s_\Im-\alpha_{l,\hat n_t}(\cos\phi_l\hat s_\Im-\sin\phi_l\hat s_\Re)\right)\right|^2.
\end{aligned}
\end{equation}
To facilitate the representation, we define the real part information $\gamma_\Re$  and imaginary part information $\gamma_\Im$ within the absolute value as
\begin{subequations}\label{cpep3n3}
\begin{align}
\gamma_\Re &= \sum\limits_{l=1}^{L}\beta_l\left(\alpha_{l,n_t}s_\Re-\alpha_{l,\hat n_t}(\cos\phi_l\hat s_\Re+\sin\phi_l\hat s_\Im)\right),\\
\gamma_\Im &= \sum\limits_{l=1}^{L}\beta_l\left(\alpha_{l,n_t}s_\Im-\alpha_{l,\hat n_t}(\cos\phi_l\hat s_\Im-\sin\phi_l\hat s_\Re)\right).
\end{align}
\end{subequations}
Under CLT, both $\gamma_\Re$ and $\gamma_\Im$ obey real Gaussian distribution. As such, it is our primary goal to determine the expectation and variance of $\gamma_\Re$ and $\gamma_\Im$.

On the one hand, since each element of RIS is independent of each other, the expectation of (\ref{cpep3n3}a) and (\ref{cpep3n3}b) can be respectively expressed as
\begin{equation}\label{cpep3n4}
\begin{aligned}
&E[\gamma_\Re] = \sum\nolimits_{l=1}^{L}E[\beta_l\left(\alpha_{l,n_t}s_\Re-\alpha_{l,\hat n_t}(\cos\phi_l\hat s_\Re+\sin\phi_l\hat s_\Im)\right)]\\
&= \sum\nolimits_{l=1}^{L}E[\beta_l]E[\alpha_{l,n_t}s_\Re-\alpha_{l,\hat n_t}(\cos\phi_l\hat s_\Re+\sin\phi_l\hat s_\Im)]\\
&=\sum\nolimits_{l=1}^{L}E[\beta_l]E[\alpha_{l,n_t}s_\Re]-E[\alpha_{l,\hat n_t}(\cos\phi_l\hat s_\Re+\sin\phi_l\hat s_\Im)]\\
&=\sum\nolimits_{l=1}^{L}E[\beta_l]E[\alpha_{l,n_t}]s_\Re-E[\alpha_{l,\hat n_t}]E[\cos\phi_l\hat s_\Re+\sin\phi_l\hat s_\Im]\\
&=\sum\nolimits_{l=1}^{L}E[\beta_l]E[\alpha_{l,n_t}]s_\Re-E[\alpha_{l,\hat n_t}]\\
&\times\left(E[\cos\phi_l]\hat s_\Re+E[\sin\phi_l]\hat s_\Im\right),
\end{aligned}
\end{equation}
\begin{equation}\label{cpep3n5}
\begin{aligned}
&E[\gamma_\Im] = \sum\nolimits_{l=1}^{L}E[\beta_l\left(\alpha_{l,n_t}s_\Im-\alpha_{l,\hat n_t}(\cos\phi_l\hat s_\Im-\sin\phi_l\hat s_\Re)\right)]\\
&=\sum\nolimits_{l=1}^{L}E[\beta_l]E[\alpha_{l,n_t}s_\Im-\alpha_{l,\hat n_t}(\cos\phi_l\hat s_\Im-\sin\phi_l\hat s_\Re)]\\
&=\sum\nolimits_{l=1}^{L}E[\beta_l]E[\alpha_{l,n_t}s_\Im]-E[\alpha_{l,\hat n_t}(\cos\phi_l\hat s_\Im-\sin\phi_l\hat s_\Re)]\\
&=\sum\nolimits_{l=1}^{L}E[\beta_l]E[\alpha_{l,n_t}]s_\Im-E[\alpha_{l,\hat n_t}]E[\cos\phi_l\hat s_\Im-\sin\phi_l\hat s_\Re]\\
&=\sum\nolimits_{l=1}^{L}E[\beta_l]E[\alpha_{l,n_t}]s_\Im-E[\alpha_{l,\hat n_t}]\\&\times\left(E[\cos\phi_l]\hat s_\Im-E[\sin\phi_l]\hat s_\Re\right).
\end{aligned}
\end{equation}
It can be observed that $\beta_l$, $\alpha_{l,n_t}$, and $\phi_l$ are the three main constituent components of (\ref{cpep3n4}) and (\ref{cpep3n5}), where $\beta_l$ and $\alpha_{l,n_t}$ denote that the magnitudes of $h_l$ and $g_{l,n_t}$ obeying the Rayleigh distribution. For this reason, we focus on finding the distribution of $\phi_l$ via the {\bf Lemma 2}.
\begin{lemma}
The PDF of variable $\phi_l$ can be expressed as
\begin{equation}
f(\phi_l)=\left\{
\begin{aligned}
 & \frac{\phi_l+2\pi}{4\pi^2},  \ \ \  \phi_l\in[-2\pi,0]      \\
 & \frac{-\phi_l+2\pi}{4\pi^2},  \ \ \  \phi_l\in[0,2\pi].
\end{aligned}
\right.
\end{equation}
\end{lemma}
\emph{ Proof:} Please refer to Appendix D.
$\hfill\blacksquare$

\begin{equation}\small\label{phic1}
\begin{aligned}
E[\cos \phi_l]\!=\!&\int_{-2\pi}^0\!\frac{\phi_l\!+\!2\pi}{4\pi^2}\cos \phi_ld\phi_l\!+\!\int_0^{2\pi}\frac{-\phi_l\!+\!2\pi}{4\pi^2}\!\cos \phi_l d\phi_l\!=\!0,
\end{aligned}
\end{equation}
\begin{equation}\small\label{phis1}
\begin{aligned}
E[\sin \phi_l]\!=\!&\int_{-2\pi}^0\frac{\phi_l\!+\!2\pi}{4\pi^2}\sin \phi_ld\phi_l\!+\!\int_0^{2\pi}\frac{-\phi_l\!+\!2\pi}{4\pi^2}\sin \phi_l d\phi_l\!=\!0.
\end{aligned}
\end{equation}
Substituting (\ref{phic1}) and (\ref{phis1}) into (\ref{cpep3n4}) and (\ref{cpep3n5}), the mean values of $\gamma_\Re$ and $\gamma_\Im$ can be simplified to
\begin{equation}\label{phis2}
E[\gamma_\Re]=\sum_{l=1}^{L}E[\beta_l\alpha_{l,n_t}]s_\Re,\ \
E[\gamma_\Im]= \sum_{l=1}^{L}E[\beta_l\alpha_{l,n_t}]s_\Im.
\end{equation}
Based on (\ref{doumean}) and CLT, (\ref{phis2}) can be further given by
\begin{equation}\label{phis3}
E[\gamma_\Re]=\frac{\pi L}{4}s_\Re,\ \ \
E[\gamma_\Im] = \frac{\pi L}{4}s_\Im.
\end{equation}

On the other hand, the variance of (\ref{cpep3n3}a) and (\ref{cpep3n3}b) can be respectively expressed as
\begin{equation}\label{phixs3}
\begin{aligned}
Var[\gamma_\Re]&=E[\gamma_\Re^2]-E^2[\gamma_\Re],
Var[\gamma_\Im]=E[\gamma_\Im^2]-E^2[\gamma_\Im].
\end{aligned}
\end{equation}
The second moment of $\gamma_\Re$ can be evaluated as
\begin{equation}\label{ere}
\begin{aligned}
&E[\gamma_\Re^2]=\sum_{l=1}^{L}E[\beta_l^2\left(\alpha_{l,n_t}s_\Re-\alpha_{l,\hat n_t}(\cos\phi_l\hat s_\Re+\sin\phi_l\hat s_\Im)\right)^2]\\
&=\sum_{l=1}^{L}E\left[\beta_l^2(\alpha_{l,n_t}^2|s_\Re|^2+\alpha_{l,\hat n_t}^2(\cos\phi_l\hat s_\Re+\sin\phi_l\hat s_\Im)^2)\right]\\&-2E[\alpha_{l,n_t}s_\Re\alpha_{l,\hat n_t}(\cos\phi_l\hat s_\Re+\sin\phi_l\hat s_\Im)].
\end{aligned}
\end{equation}
Based on (\ref{phic1}) and (\ref{phis1}), the (\ref{ere}) can be simplified to
\begin{equation}\label{ere1}
\begin{aligned}
&E[\gamma_\Re^2]
\!=\!\sum_{l=1}^{L}\!E\left[\beta_l^2\left(\alpha_{l,n_t}^2|s_\Re|^2\!+\!\alpha_{l,\hat n_t}^2(\cos\phi_l\hat s_\Re\!+\!\sin\phi_l\hat s_\Im)^2\right)\right]\\
&=\sum_{l=1}^{L}E\left[\beta_l^2(\alpha_{l,n_t}^2|s_\Re|^2\!+\!\alpha_{l,\hat n_t}^2(\cos^2\phi_l|\hat s_\Re|^2+\sin^2\phi_l|\hat s_\Im|^2\right.\\&\left.+2\cos\phi_l\hat s_\Re\sin\phi_l\hat s_\Im))\right].
\end{aligned}
\end{equation}
By means of (\ref{phic1}) and (\ref{phis1}), we can further obtain
\begin{small}
\begin{equation}
\begin{aligned}
&E[\gamma_\Re^2]
\!=\!\sum_{l=1}^{L}E\!\left[\beta_l^2(\alpha_{l,n_t}^2|s_\Re|^2\!+\!\alpha_{l,\hat n_t}^2(\cos^2\phi_l |\hat s_\Re|^2\!+\!\sin^2\phi_l|\hat s_\Im|^2))\right]\\
&=\sum\nolimits_{l=1}^{L}E[\beta_l^2(\alpha_{l,n_t}^2|s_\Re|^2+\alpha_{l,\hat n_t}^2\left((1+\cos2\phi_l)|\hat s_\Re|^2/{2}\right.\\&\left.+(1-\cos2\phi_l)|\hat s_\Im|^2/{2}\right))].
\end{aligned}
\end{equation}
\end{small}%
After some calculations, we have
\begin{equation}\label{ere2}
\begin{aligned}
E[\gamma_\Re^2]
&=\sum\nolimits_{l=1}^{L}E[\beta_l^2(\alpha_{l,n_t}^2|s_\Re|^2+\alpha_{l,\hat n_t}^2\frac{|\hat s|^2}{2})].
\end{aligned}
\end{equation}
Since $\beta_l$, $\alpha_{l,n_t}$, and $\alpha_{l,\hat n_t}$ follow Rayleigh distribution, we can get
\begin{equation}\label{ere3}
\begin{aligned}
&E[\beta_l^2]=E^2[\beta_l]+Var[\beta_l]=\frac{\pi}{4}+\frac{4-\pi}{4}=1,\\
&E[\alpha_{l,n_t}^2]=E[\alpha_{l,\hat n_t}^2]=1.
\end{aligned}
\end{equation}
By replacing (\ref{ere3}) in (\ref{ere2}) and applying the CLT, we can recast $E[\gamma_\Re^2]$ as
\begin{equation}\label{ere4}
\begin{aligned}
E[\gamma_\Re^2]=
|s_\Re|^2L+\frac{|\hat s|^2L}{2}.
\end{aligned}
\end{equation}
Combining (\ref{ere4}), (\ref{phis3}), and (\ref{phixs3}), we can calculate the variance of $\gamma_\Re$ as
\begin{equation}
Var[\gamma_\Re]
=\frac{(16-\pi^2)|s_\Re|^2L}{16}+\frac{|\hat s|^2L}{2}.
\end{equation}

\begin{theorem}
The (\ref{phixs3}) can be given by
\begin{equation}
Var[\gamma_\Im]
=\frac{(16-\pi^2)|s_\Im|^2L}{16}+\frac{|\hat s|^2L}{2}.
\end{equation}
\end{theorem}
\emph{Proof:} Please refer to Appendix E.
$\hfill\blacksquare$

After that, we address the covariance of $\gamma_\Re\gamma_\Im$ as
\begin{equation}\label{covdx1}
Cov[\gamma_\Re\gamma_\Im]=E[\gamma_\Re\gamma_\Im]-E[\gamma_\Re]E[\gamma_\Im].
\end{equation}
Since $E[\gamma_\Re]$ and $E[\gamma_\Im]$ are obtained via (\ref{phis3}), we derive the $E[\gamma_\Re\gamma_\Im]$ as follows:
\begin{equation}\label{covd1}
\begin{aligned}
&E[\gamma_\Re\gamma_\Im]
= \sum_{l=1}^{L}E\left[\beta_l^2\left(\alpha_{l,n_t}s_\Re-\alpha_{l,\hat n_t}(\cos\phi_l\hat s_\Re+\sin\phi_l\hat s_\Im)\right)\right.\\&\left.\times\left(\alpha_{l,n_t}s_\Im-\alpha_{l,\hat n_t}(\cos\phi_l\hat s_\Im-\sin\phi_l\hat s_\Re)\right) \right].
\end{aligned}
\end{equation}
By utilizing the (\ref{ere3}), we can further calculated (\ref{covd1}) as
\begin{equation}
\begin{aligned}
&E[\gamma_\Re\gamma_\Im]
= \sum_{l=1}^{L}E\left[\left(\alpha_{l,n_t}s_\Re-\alpha_{l,\hat n_t}(\cos\phi_l\hat s_\Re+\sin\phi_l\hat s_\Im)\right)\right.\\&\times\left.\left(\alpha_{l,n_t}s_\Im-\alpha_{l,\hat n_t}(\cos\phi_l\hat s_\Im-\sin\phi_l\hat s_\Re)\right) \right]\\
&= \sum\nolimits_{l=1}^{L}E\left[\alpha_{l,n_t}^2s_\Re s_\Im +\alpha^2_{l,\hat n_t}(\cos\phi_l\hat s_\Re+\sin\phi_l\hat s_\Im)\right.\\&\times\left.(\cos\phi_l\hat s_\Im\!-\!\sin\phi_l\hat s_\Re)
\!-\!\alpha_{l,n_t}s_\Re\alpha_{l,\hat n_t}(\cos\phi_l\hat s_\Im\!-\!\sin\phi_l\hat s_\Re)\right.\\&\left.
-\alpha_{l,n_t}s_\Im\alpha_{l,\hat n_t}(\cos\phi_l\hat s_\Re+\sin\phi_l\hat s_\Im)\right].
\end{aligned}
\end{equation}
Resort to (\ref{phic1}), (\ref{phis1}), and (\ref{ere3}), we can rewrite $E[\gamma_\Re\gamma_\Im]$ as
\begin{equation}\label{covd2}
\begin{aligned}
&E[\gamma_\Re\gamma_\Im]
= \sum_{l=1}^{L}E\left[\alpha_{l,n_t}^2s_\Re s_\Im +\alpha^2_{l,\hat n_t}(\cos\phi_l\hat s_\Re+\sin\phi_l\hat s_\Im)\right.\\&\times\left.(\cos\phi_l\hat s_\Im-\sin\phi_l\hat s_\Re)\right]\\
&= \sum_{l=1}^{L}E\left[s_\Re s_\Im \!+\!(\cos\phi_l\hat s_\Re+\sin\phi_l\hat s_\Im)(\cos\phi_l\hat s_\Im\!-\!\sin\phi_l\hat s_\Re)\right].
\end{aligned}
\end{equation}
Based on (\ref{phic1}) and (\ref{phis1}), the (\ref{covd2}) can be evaluated as
\begin{equation}
\begin{aligned}
& E[\gamma_\Re\gamma_\Im]
=\sum_{l=1}^{L}E\left[s_\Re s_\Im +\cos^2\phi_l\hat s_\Re\hat s_\Im-\sin^2\phi_l\hat s_\Re\hat s_\Im\right.\\&\left.
-\cos\phi_l\hat s_\Re\sin\phi_l\hat s_\Re+\sin\phi_l\hat s_\Im\cos\phi_l\hat s_\Im\right]\\
&= \sum_{l=1}^{L}E\left[s_\Re s_\Im +\cos^2\phi_l\hat s_\Re\hat s_\Im-\sin^2\phi_l\hat s_\Re\hat s_\Im
\right]\\
&= \sum_{l=1}^{L}E[s_\Re s_\Im +\frac{1+\cos2\phi_l}{2}\hat s_\Re\hat s_\Im-\frac{1-\cos2\phi_l}{2}\hat s_\Re\hat s_\Im
].
\end{aligned}
\end{equation}
After some mathematical operations, we have
\begin{equation}\label{rrri}
\begin{aligned}
E[\gamma_\Re\gamma_\Im]
= \sum\nolimits_{l=1}^{L}E\left[s_\Re s_\Im
\right]= s_\Re s_\Im L.
\end{aligned}
\end{equation}
Combining (\ref{covdx1}), (\ref{rrri}), and (\ref{phis3}), we obtain
\begin{equation}
Cov[\gamma_\Re\gamma_\Im]=s_\Re s_\Im L-\frac{\pi^2s_\Re s_\Im L}{16}=\frac{(16-\pi^2)s_\Re s_\Im L}{16}.
\end{equation}

In this manner, let us define $\mathbf{\Gamma}=\mathbf{z}^T\mathbf{Az}$, where $\mathbf{z} = [\gamma_\Re \ \ \gamma_\Im]^T$ and $\mathbf{A}=\mathbf{I}_2$.
As a result, the mean vector and covariance matrix of $\mathbf{z}$ can be described as
\begin{equation}
\boldsymbol{\mu}=
\begin{bmatrix}
\frac{L\pi}{4}s_\Re & \frac{L\pi}{4}s_\Im
\end{bmatrix},
\end{equation}
\begin{equation}
\mathbf{V}=
\begin{bmatrix}
Var[\gamma_\Re] & Cov[\gamma_\Re\gamma_\Im]\\
Cov[\gamma_\Im\gamma_\Re] & Var[\gamma_\Im]
\end{bmatrix}.
\end{equation}
According to \cite{mathai1992}, the MGF of $\Gamma$ can be calculated by
\begin{equation}\label{nmgf}
\begin{aligned}
&{\rm MGF}_\Gamma(x)=\left({\rm det}(\mathbf{I}-2x\mathbf{AV})\right)^{-\frac{1}{2}}\\&\times\exp(
-\frac{1}{2}\boldsymbol{\mu}^H\left(\mathbf{I}-(\mathbf{I}-2x\mathbf{AV})^{-1}\right)\mathbf{V}^{-1}\boldsymbol{\mu}
).
\end{aligned}
\end{equation}
In this stage, we consider (\ref{cpep3n1}) and (\ref{nmgf}).
Hence, the UPEP can be given by
\begin{equation}
\begin{aligned}
&\bar P_b
=\frac{1}{\pi}\int_0^{\frac{\pi}{2}}({\rm det}(\mathbf{I}+\frac{\rho}{2\sin^2\theta}\mathbf{AV}))^{-\frac{1}{2}}\\&\times\exp(
-\frac{1}{2}\boldsymbol{\mu}^H(\mathbf{I}-(\mathbf{I}+\frac{\rho}{2\sin^2\theta}\mathbf{AV})^{-1})\mathbf{V}^{-1}\boldsymbol{\mu}
) d\theta.
\end{aligned}
\end{equation}
To obtain closed-form expression of UPEP, we deal with them with the GCQ method.
We let $\theta=\frac{\pi}{4}\varpi+\frac{\pi}{4}$, after some calculations, the UPEP in this case can be written as
\begin{equation}
\begin{aligned}
&\bar P_b
=\frac{\pi}{4Q}\sum\nolimits_{q=1}^Q ({\rm det}(\mathbf{I}+\frac{\rho}{2\sin^2\left(\frac{\pi}{4}\varpi+\frac{\pi}{4}\right)}\mathbf{AV}))^{-\frac{1}{2}}\\
&\times\exp(
-\frac{1}{2}\boldsymbol{\mu}^H(\mathbf{I}-(\mathbf{I}+\frac{\rho}{2\sin^2\left(\frac{\pi}{4}\varpi+\frac{\pi}{4}\right)}\mathbf{AV})^{-1})\mathbf{V}^{-1}\boldsymbol{\mu}
)\\&+R_Q,
\end{aligned}
\end{equation}
where $\varpi=\cos\left(\frac{2q-1}{2Q}\pi\right)$ and $R_Q$ denotes the error term.
\subsection{ABEP}
In general, there is no precise ABEP expression for RIS-SM systems for arbitrary spatial and symbol domain modulation orders.
Consequently, the union upper bound technique is adopted to derive the tight upper bound of ABEP on the RIS-SM system. Given this, the expression of ABEP can be characterized as
\begin{equation}\label{abep}
{\rm ABEP} \leq \sum_{n_t = 1}^{N_t}\sum_{m=1}^M\sum_{\hat n_t = 1}^{N_t}\sum_{\hat m=1}^M\frac{\bar P_b N([n_t, m]\to [\hat n_t, \hat m])}{N_tM\log_2(N_tM)},
\end{equation}
where $N([n_t, m]\to [\hat n_t, \hat m])$ denotes the Hamming distance between the transmitted signal index $n_t$ and $s$ and detected signal $\hat n_t$ and $\hat s$.
It is worth noting that the equation sign holds if and only if $N_t=2$ and $M=1$.
\subsection{Ergodic Capacity Analysis}
It is worth mentioning that the Shannon capacity gives the upper limit of ergodic capacity in terms of the Gaussian distribution input. However, in real communication systems, the signal does not always follow the Gaussian distribution. In the proposed RIS-SM system, the spatial and symbol domain signals of the finite inputs are used to transmit information. Consequently, the ergodic capacity of the RIS-SM via the continuous output memoryless channel (DCMC) can be given as \cite{an2015mutual}
\begin{equation}\label{EC1}
EC={E}[\max\limits_{n_t\in\mathcal{N}, s\in\mathcal{S}}\mathcal{I}\left(G_{n_t},s;y\right)].
\end{equation}
Without loss of generality, the mutual information can be given via the chain rule as
\begin{equation}\label{EC2}
\mathcal{I}\left(G_{n_t},s;{y}\right) = \mathcal{H}({y}) - \mathcal{H}\left({y} |G_{n_t},s\right),
\end{equation}
where $\mathcal{H}({y})$ and $\mathcal{H}({y} |G_{n_t},s)$ denotes the differential entropies of the received signal and noise, respectively.
In particular, the term of $\mathcal{H}\left({y} |G_{n_t},s\right)$ can be written as
\begin{equation}\label{EC3}
\mathcal{H}\left({y} |G_{n_t},s\right) = \log_2\left(\pi\exp(1)N_0\right).
\end{equation}
It should be noted that maximizing (\ref{EC1}) is equivalent to maximizing the term of $\mathcal{H}({y})$ as
\begin{equation}\label{EC4}
\begin{aligned}
\mathcal{H}({y})&=\int f(y)\log_2(f(y))dy\overset{(a)}{\geq} -\log_2\left(\int \left(f(y)\right)^2dy\right),
\end{aligned}
\end{equation}
where $(a)$ stands for the use of Jensen's Inequality to obtain the lower bound of $\mathcal{H}({y})$.
According to \cite{soon2006on}, it is known that $\mathcal{H}({y})$ can obtain the maximum value when the discrete input signal obeys an equal probability distribution.
As a consequence, the $f(y)$ can be expressed as
\begin{equation}\label{EC5}
\begin{aligned}
f(y) &= \frac{1}{N_tM\pi N_0}\sum_{n_t=1}^{N_t}\sum_{m=1}^Mf\left(y|G_{n_t},s\right)\\
&=\frac{1}{N_tM}\sum_{n_t=1}^{N_t}\sum_{m=1}^M\exp({-|y-\sqrt{P_t}G_{n_t} s|^2}/{N_0}).
\end{aligned}
\end{equation}
Subsequently, the (\ref{EC2}) can be written via (\ref{EC3}) and (\ref{EC4})  as
\begin{equation}\label{EC6}
\begin{aligned}
&\mathcal{I}\left(G_{n_t},s;{y}\right) \approx 2\log_2(N_tM)\\
&-\log_2\!(N_tM\!+\!\sum_{n_t=1}^{N_t}\!\sum_{m=1}^{M}\!\sum_{\hat n_t\neq n_t}^{N_t}\!\sum_{\hat m\neq m}^{M}\!\exp(\frac{-\rho\left|G_{n_t}s\!-\!G_{\hat n_t}\hat s\right|^2}{2})\!).
\end{aligned}
\end{equation}
Afterwards, we evaluate the (\ref{EC6}) over fading channel as
\begin{equation}\label{EC7}
\begin{aligned}
&\mathcal{I}\left(G_{n_t},s;{y}\right) \approx 2\log_2(N_tM)\!-\!\log_2\left(N_tM\!
\right.\\
&\left.+\!\sum_{n_t=1}^{N_t}\sum_{m=1}^{M}\sum_{\hat n_t\neq n_t}^{N_t}\sum_{\hat m\neq m}^{M}\underbrace{E\left[\exp\left(\frac{-\rho\left|G_{n_t}s\!-\!G_{\hat n_t}\hat s\right|^2}{2}\right)\right]}_{\Lambda}\right),
\end{aligned}
\end{equation}
where $E[\exp({\rho\left|G_{n_t}s\!-\!G_{\hat n_t}\hat s\right|^2}/{2})]$ denotes the MGF of $G_{\hat n_t}$.
Subsequently, we refer to the derivation process of ABEP to obtain the $\Lambda$ as
\begin{equation}\label{ecmgf2}
\begin{aligned}
\Lambda=&\left({\rm det}(\mathbf{I}-\rho\mathbf{AV})\right)^{-\frac{1}{2}}\\&\times\exp\left(
-\frac{1}{2}\boldsymbol{\mu}^H\left(\mathbf{I}-(\mathbf{I}-\rho\mathbf{AV})^{-1}\right)\mathbf{V}^{-1}\boldsymbol{\mu}
\right).
\end{aligned}
\end{equation}
By substituting (\ref{ecmgf2}) into (\ref{EC7}), we can obtain the closed-form expression on the ergodic capacity of the RIS-SM system.

\section{Simulation and Analytical Results}
In this section, we first address the complexity of the three detectors presented in this paper. Then, the performance of the proposed RIS-SM scheme is evaluated in terms of ABEP and ergodic capacity.
\subsection{Complexity Analysis}

In this section, we characterize the computational complexity by the number of operations of real multiplications and real additions. For convenience, we denote the complexities of real number multiplication and addition as $A$ and $B$, respectively. Then, the complexities of complex addition and multiplication are recorded as $2B$ and $4A+2B$, respectively. Besides, the complexity of the real number root operation is $A$. Furthermore, the absolute value square operation of complex numbers requires a computational complexity of $2A+B$.
After that, we analyze the detection complexity of ML, TSML, and GD detectors of the proposed RIS-SM scheme, respectively.

\begin{figure}[t]
 \centering
 \subfigure[Flops versus $L$]
 {
  \begin{minipage}[b]{0.22\textwidth}
   \centering
   \includegraphics[width=4cm]{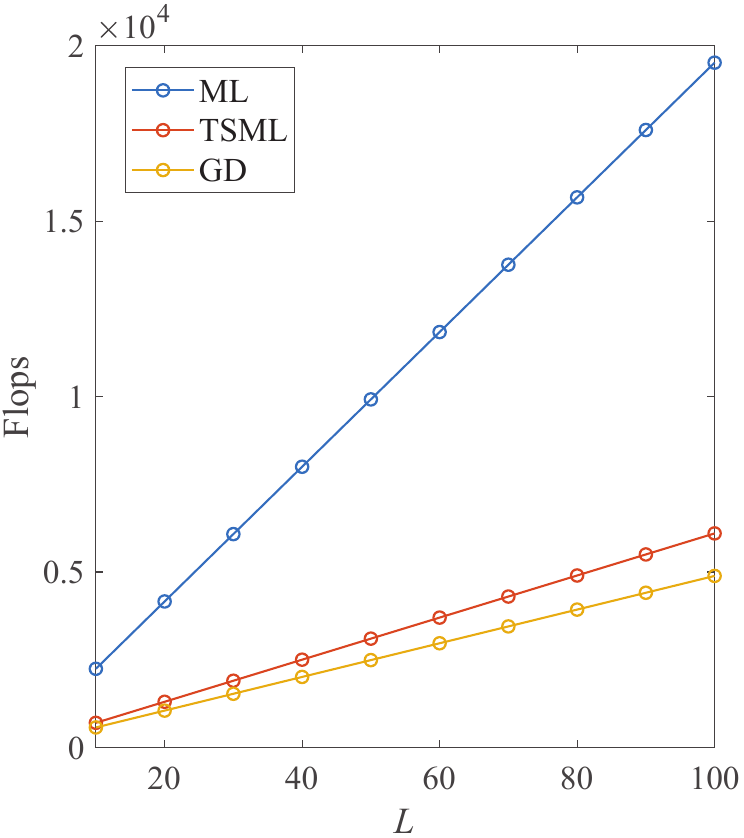}
  \end{minipage}
 }
 \subfigure[Flops versus $N_t$]
    {
     \begin{minipage}[b]{0.20\textwidth}
      \centering
      \includegraphics[width=3.7cm]{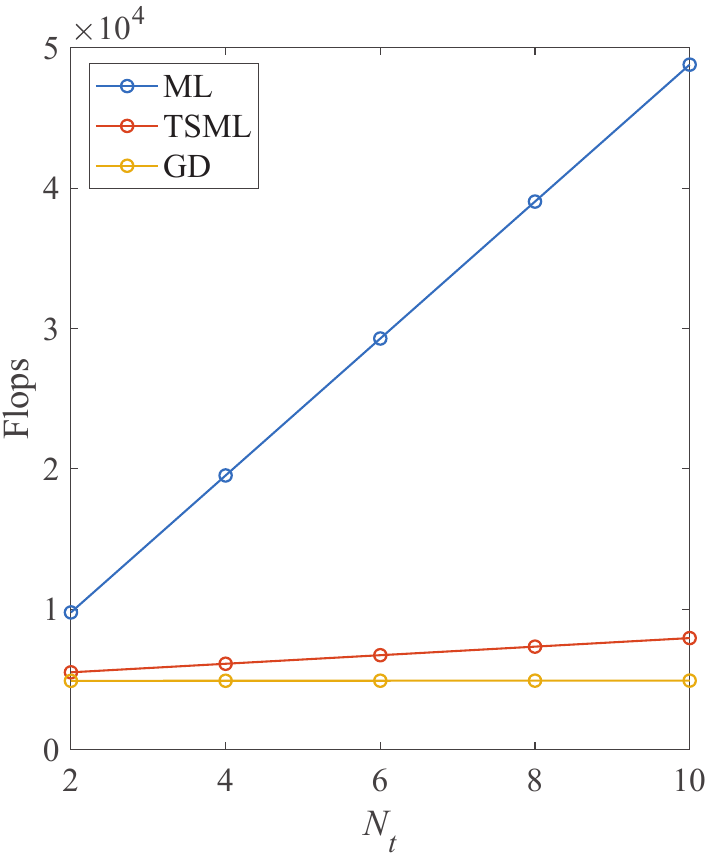}
     \end{minipage}
    }
  \subfigure[Flops versus $M$]
    {
     \begin{minipage}[b]{0.22\textwidth}
      \centering
      \includegraphics[width=4cm]{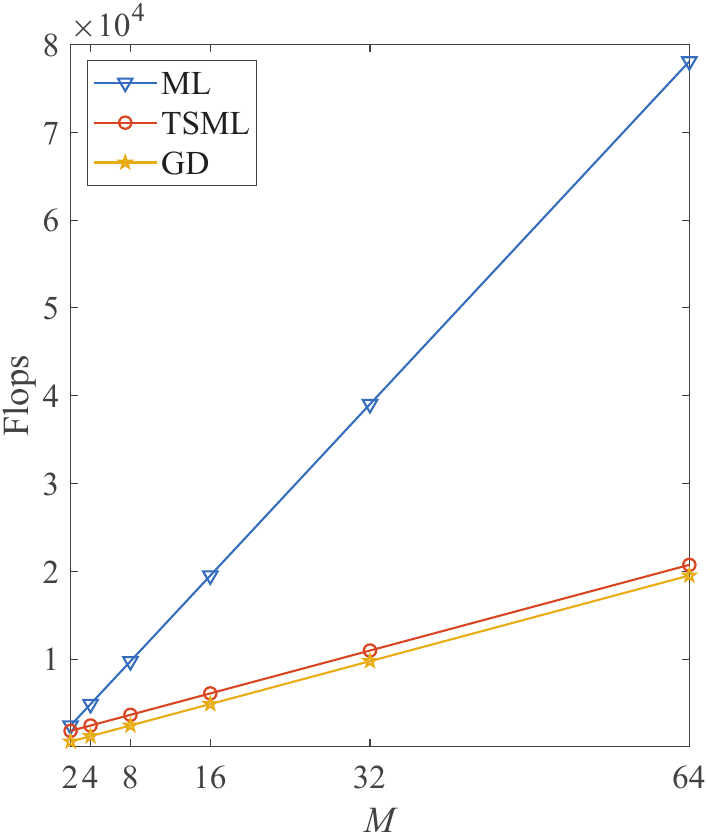}
     \end{minipage}
    }
\caption{\small{Complexity analysis of different detectors.}}
\label{complextiyana}
\end{figure}
\begin{table}[t]
\centering
\caption{\small{Detector Complexity}}
\begin{tabular}{|c|c|c|}
\hline {\bf Detector} & {\bf Multiplication} & {\bf Summation} \\
\hline ML & $(3L+5)N_tM$ & $(2L+1)N_tM$ \\
\hline TSML & $(3L+5)(N_t+M)$ & $(2L+1)(N_t+M)$ \\
\hline GD & $2N_t+(3L+5)M$ & $N_t+(2L+1)M$ \\
\hline
\end{tabular}
\end{table}

For ML detector, the complexity of $\alpha_{l,n_t}\beta_l s$ requires $3A$, then $\sum\nolimits_{l=1}^{L}\alpha_{l,n_t}\beta_l s$ can be calculated by $3AL+2(L-1)B$. Also, $\sqrt{P_t}$ provides one multiplication $A$. Based on this, $\sqrt{P_t}\sum\nolimits_{l=1}^{L}\alpha_{l,n_t}\beta_l s$ requires $(3L+3)A+2(L-1)B$. Taking this into account, the obtained value can be calculated by subtracting from $y$, the corresponding complexity can be obtained as $(3L+3)A+2LB$. In addition, the term of $|y-\sqrt{P_t}\sum\nolimits_{l=1}^{L}\alpha_{l,n_t}\beta_l s|^2$ requires the calculation complexity denoted by $(3L+5)A+(2L+1)B$.
Subsequently, we obtain the computational complexity of the ML detector is  $(3L+5)N_tMA+(2L+1)N_tMB$ via an exhaustive search method. For the TSML detector, the computation complexity of (\ref{tsml}) can be given by $(3L+5)N_tA+(2L+1)N_tB$, which can be referenced by the process of ML detector. Similarly, the term of (\ref{sicsymbol}) can be computed by $(3L+5)MA+(2L+1)MB$. As such, the complexity of the TSML detector is $(3L+5)(N_t+M)A (2L+1)(N_t+M)B$. For the GD detector, the $|y|^2$ operation requires two additions and one multiplication. Then, the transmit antenna index is recovered by employing the ergodic search approach. As such, the complexity of the GD detector can be given as $(2N_t (3L+5)M)A+(N_t+(2L+1)M)B$. Table I exhibits the computational complexity expressions of ML, TSML, and GD detectors.

\begin{figure}[t]
  \centering
  \includegraphics[width=8cm]{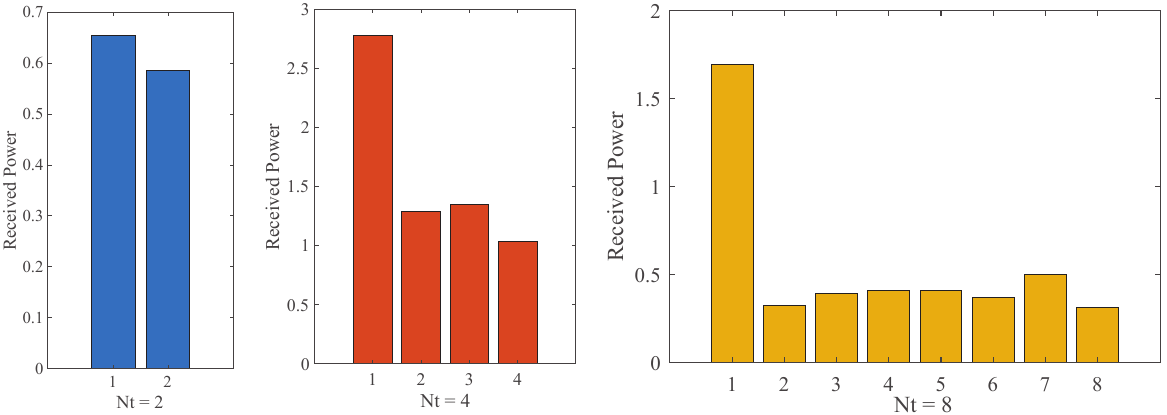}\\
  \caption{The received power with respect to different transmit antennas.}\label{GDpower}
\end{figure}

Fig. \ref{complextiyana} provides the decoding computational complexity of different detectors under different parameters of the proposed RIS-SM system.
It is evident from Fig. \ref{complextiyana} that the complexity of ML is always the highest, whether with the number of reflection units, the number of transmit antennas, or the modulation order of the signal. This is due to the fact that the ML algorithm traverses the two-dimensional signal space of $M$ and $N_t$.
Moreover, we can observe that the TSML detector can considerably decrease the calculation complexity.
This is because the TSML algorithm divides the spatial and symbol domain information into two sequential parts for decoding and detection. Specifically, the spatial domain information is first detected using the ML algorithm, and the ML algorithm again recovers the symbol domain information.
In addition, the GD detector can achieve the lowest calculation complexity compared with ML and TSML detectors.
This is because the phase shift of the RIS is regulated to be the optimal phase shift from the transmit antenna to the receive antenna so that only the transmit antenna signal reflected by the RIS reaches the receive antenna with the maximum energy. On the contrary, the remaining antenna signals of the BS with RIS reach the UE with less energy.
Considering this property, we designed the GD detector via the received energy intensity.
When the number of antennas at the BS side is 2, 4, and 8, we plot the corresponding result Fig. \ref{GDpower}, where the first antenna is activated.
As expected, the detection of the first transmit antenna is somewhat higher than the detection of the other transmit antennas, which provides a rationale for the GD algorithm.

\begin{figure}[t]
  \centering
  \includegraphics[width=8cm]{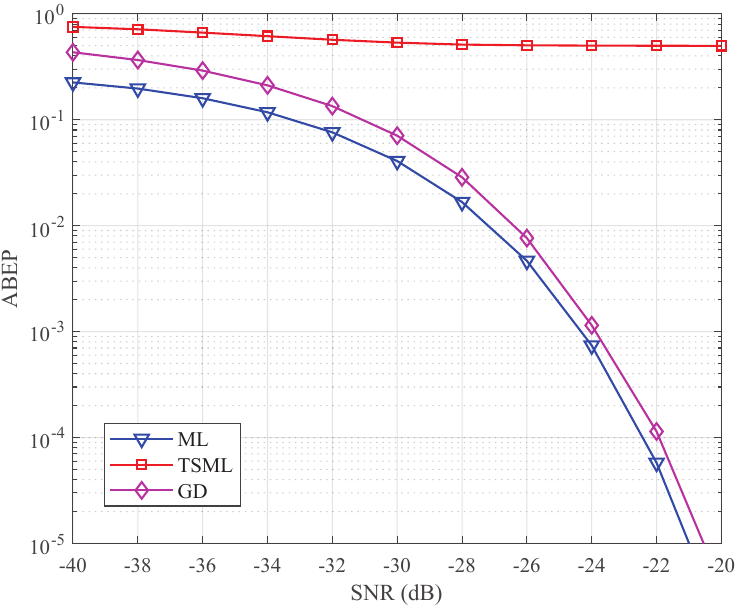}\\
  \caption{Comparison of ABEP performance with different detectors.}\label{berdet}
\end{figure}
\begin{figure}[t]
  \centering
  \includegraphics[width=8cm]{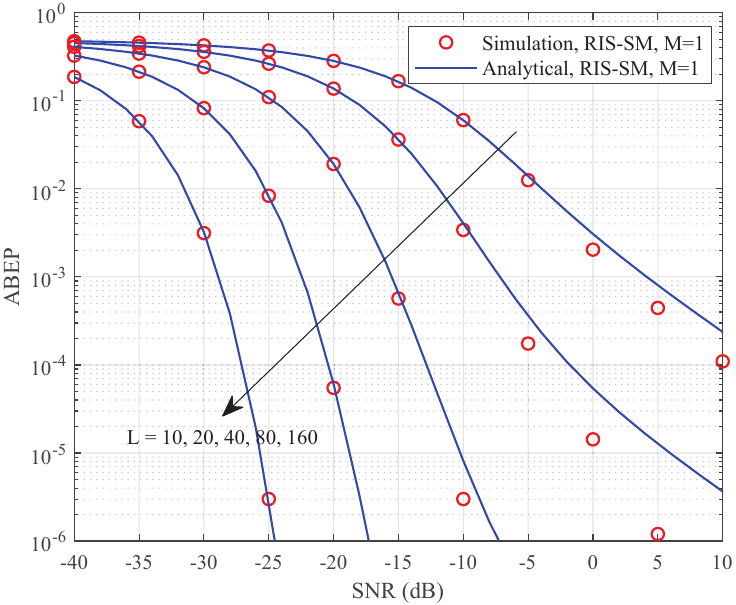}\\
  \caption{Verification of the CLT-based analytical ABEP results.}\label{CLT}
\end{figure}

\subsection{ABEP}

In Fig. \ref{berdet}, we illustrate the ABEP performance of the three detection algorithms of the proposed RIS-SM scheme with $M=2$ and $N_t=2$. From Fig. \ref{berdet}, the following observations are made. Firstly, we observe that the TSML detection algorithm is failing. Specifically, the ABEP curve decreases only slightly as the SNR increases, but this result cannot recover the original signal. The reason for this phenomenon is that when decoding spatial domain signals, the demodulation of the spatial domain signal by the symbol domain signal is equivalent to multiplicative interference. In the case of  $\tilde s=s$ on (\ref{tsml}), the corresponding probability is $\frac{1}{M}$. At this point, the decoding signal is only to recover the antenna index $n_t$ without interference signal. As the SNR increases, the value of ABEP is small. In the case of  $\tilde s\neq s$ on (\ref{tsml}), the corresponding probability is $\frac{M-1}{M}$. As the SNR increases, the ABEP value hardly changes much, so the original signal cannot be recovered and the value of ABEP is large. Combining these two cases, we can obtain that the value of ABEP is large and hardly changes with the change of SNR. Secondly, as expected, the ML algorithm obtains the optimal ABEP performance, which corresponds to the highest decoding complexity of ML. Finally, we notice that the ABEP performance obtained based on the GD algorithm is very close to the ABEP value obtained with ML. In other words, the GD algorithm can still achieve considerable ABEP performance at a very low calculation complexity.

\begin{figure}[t]
  \centering
  \includegraphics[width=8cm]{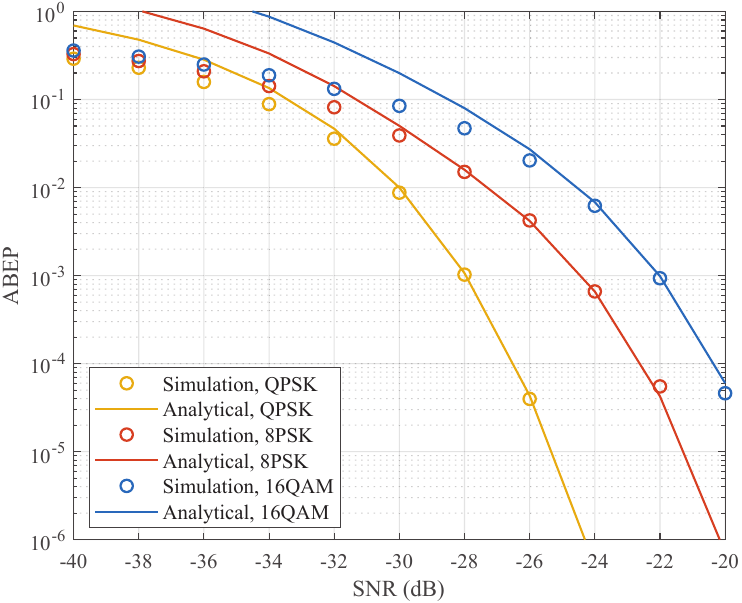}\\
  \caption{ABEP performance under different modulation orders of symbol domain.}\label{PSKQAM}
\end{figure}

\begin{figure}[t]
  \centering
  \includegraphics[width=8cm]{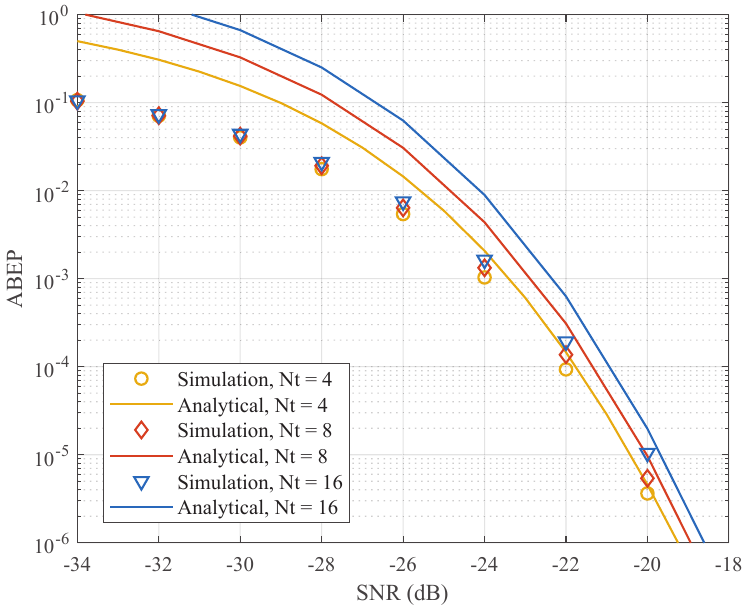}\\
  \caption{ABEP performance with different transmit antennas.}\label{NtBER}
\end{figure}

In Fig. \ref{CLT}, we validate the theoretical ABEP results obtained with the CLT approach from the Monte Carlo method. To visualize the CLT fit more closely, we set the number of transmit antennas and the symbol-domain modulation order to 2 and 1, respectively. Note that the rationale for this setting originates from the (\ref{abep}). It can be seen from Fig. \ref{CLT} that the theoretical and simulation results perfectly converge together as the number of reflective elements of RIS is 80 and 160. Whereas, the gap between the two becomes larger in case the number of elements of RIS is 10, 20, and 40. This is because the composite channel distribution more accurately approaches the Gaussian distribution when the number of reflecting elements $L$ is not less than 80. In particular, the simulation and theoretical derivations match in the low SNR region as the $L$ value is relatively small. In contrast, the gap between the two gradually becomes larger in the high SNR region. This is a result of the fact that the gap between the combined channel and the Gaussian distribution is amplified as the SNR increases.

Fig. \ref{PSKQAM}, we depict the ABEP performance versus SNR results under different modulation orders for the RIS-SM system, where the number of transmit antennas and the number of reflecting elements of the RIS are set to 2 and 100, respectively. As seen from Fig. \ref{PSKQAM}, in the low SNR region, simulation results and analytical curves do not converge. However, as the SNR increases, the two results gradually become consistent and tend to keep in agreement. This is because the analytical results in (\ref{abep}) are the tight union upper bound of ABEP rather than the exact ABEP results. In addition, we observe that the performance of the RIS-SM system deteriorates accordingly the higher the modulation order in the symbol domain. This is because we normalize the energy of the transmitted symbol $s$. When the modulation order is higher, the Euclidean distance between adjacent constellation points becomes smaller, resulting in a greater probability of decision failure.

In Fig. \ref{NtBER}, we provide the ABEP performance curves and Monte Carlo simulation results for the proposed RIS-SM system. The modulation order and the number of reflecting units of the RIS are given as 2 and 100, respectively. From the given results, it can be seen that the simulation results and the analytical results tend to match as the SNR increases. Moreover, we observe that the ABEP performance of the RIS-SM system does not degrade significantly when the number of transmit antennas increases from 4 to 16. This phenomenon illustrates that the data rate can be improved from the spatial domain. Compared to the increasing modulation order in Fig. \ref{PSKQAM}, the effect on the ABEP performance is smaller.

\begin{figure}[t]
  \centering
  \includegraphics[width=8cm]{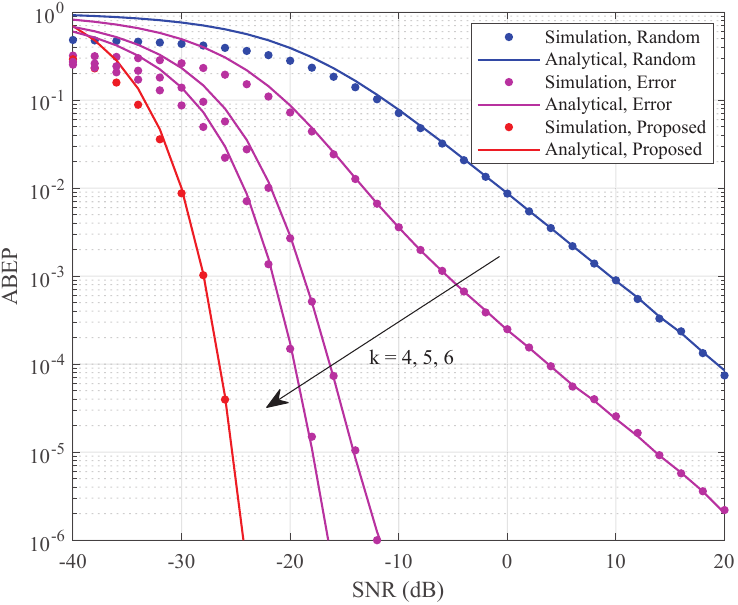}\\
  \caption{The impact of phase adjustment accuracy of RIS on ABEP performance.}\label{error}
\end{figure}

In Fig. \ref{error}, we investigate the ABEP performance for the RIS-SM scheme under phase adjustment accuracy of RIS, where the parameters are set as $M=2, N_t = 2$, and $L=100$.
Specifically, the blue curve in Fig. \ref{error} denotes the reflecting shifts that are randomly generated, and the purple curves indicate the case where there are adjustment errors in the reflecting phase shift.
For the sake of simplicity, we assume that the interval of the phase shift adjustment error is uniformly distributed between $\left[-\pi/k, \pi/k\right]$. The given results show that the theoretical curves match the simulation results in the moderate and high SNR regions. Also, the performance of the RIS-SM scheme improves with increasing phase regulation accuracy. It is worth noting that the ABEP performance of the RIS-SM system is higher than the proposed scheme for both the random phase reflection and the presence of reflecting error cases, which align with our expected results.

\begin{figure}[t]
  \centering
  \includegraphics[width=8cm]{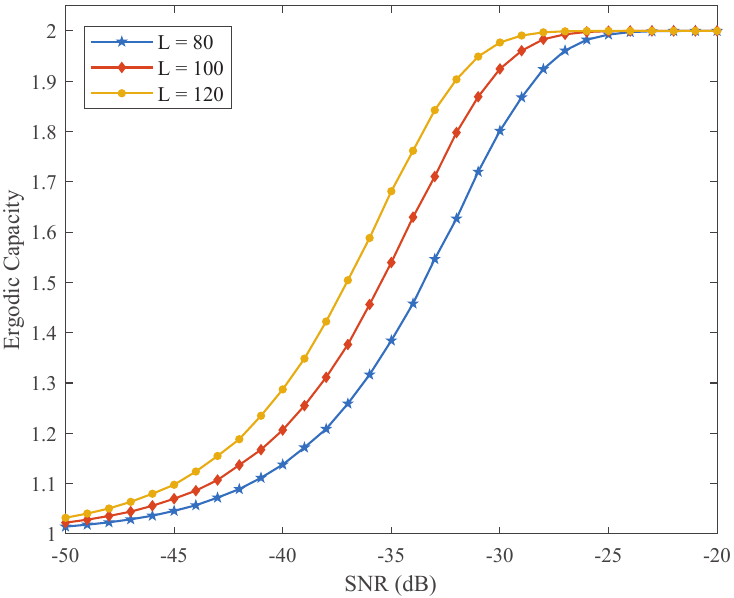}\\
  \caption{Ergodic capacity versus SNR for different $L$.}\label{ECL}
\end{figure}
\begin{figure}[t]
  \centering
  \includegraphics[width=8cm]{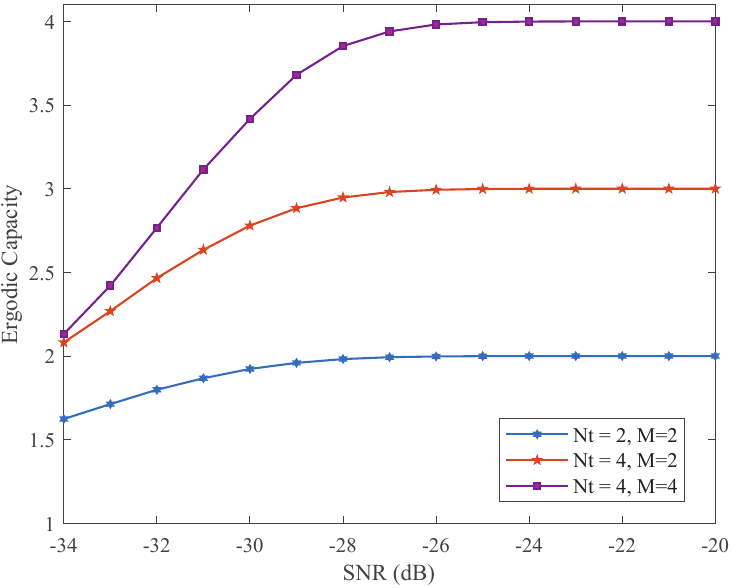}\\
  \caption{Ergodic capacity versus SNR for different $N_t$ and $M$.}\label{ECMNt}
\end{figure}
\subsection{Ergodic Capacity}
Fig. \ref{ECL} describes the analytical results obtained for the ergodic capacity of the RIS-SM system under different numbers of reflection elements. The parameters for the number of transmit antennas and the modulation order in the symbol domain are given as $N_t=2$ and $M=2$, respectively. We observe from Fig. \ref{ECMNt} that the value of EC undergoes the whole process from a sharp increase to a gradual saturation with the increase of SNR. It is worth noting that when $L$ takes different values, ergodic capacity achieves the same result under high SNR, which means that increasing the number of reflective units in RIS cannot improve the ergodic capacity of the system. Nevertheless, we observe that increasing the number of units in the RIS can speed up the EC reaching the saturation value. Specifically, when SNR = -35 dB, the EC value of $L=120$ is 0.13 and 0.3 higher than $L=100$ and $L=80$, respectively.

In Fig. \ref{ECMNt}, we exhibit the ergodic capacity versus SNR curves of the proposed RIS-SM scheme in terms of $N_t$ and $M$, where the number of reflecting elements is set as $L=100$. It is worth mentioning that the analytical curves in Fig. \ref{ECMNt} for RIS-SM systems are plotted based on (\ref{EC7}). From Fig. \ref{ECMNt}, it can be seen that saturation values of ergodic capacity are reached with increasing SNR. Specifically, the value of EC can be taken as 2 in case we use BPSK and two transmit antennas, which is consistent with our expected data rate of 2 bits per channel use (bpcu). In addition, we can improve the EC of the RIS-SM system by increasing the number of antennas in the spatial domain from $N_t=2$ to $N_t=4$ or the modulation order in the symbol domain from $M=2$ to $M=4$.

\section{Conclusion}
In this paper, we studied the RIS-assisted SM downlink communication system, proposing three detection methods. Leveraging the ML detector with the CLT, we derive the PDF of the combined channel and establish closed-form expressions for the ABEP and ergodic capacity. Monte Carlo simulations confirm the theoretical results, indicating that the GD method offers the lowest complexity with near-ML performance. The TSML detector's inability to recover signals is also noted. It is worth mentioning that the impact of RIS phase adjustment errors on ABEP performance was also studied. We further analyzed the impact of RIS phase errors on ABEP and studied the effects of system parameters on ABEP and ergodic capacity, providing insights for system optimization.
In future work, the impact of antenna selection techniques on RIS-SM schemes can be investigated. Also, it can be studied from the point of view of multiple RIS or multiple receive antennas. Furthermore, the movable antenna \cite{zhou2024movable,he2024mob} on the proposed system could be investigated.

\begin{appendices}
\section{Proof of Theorem 1}
It is worth noting that $\alpha_{l,n_t}\beta_l, 1\leq l\leq L$ is independently and identically distributed (i.i.d.), so that the sum $\xi$ is also i.i.d. random, which prompts the utilization of the CLT.

According to \cite{sag2014on}, the PDF of $X=|\alpha_{l,n_t}\beta_l|^2$ can be evaluated as
\begin{equation}
f_X(x) = 2K_0(2\sqrt{x}),
\end{equation}
where $K_0(\cdot)$ denotes the zeroth order modified Bessel function of the second kind.

Without loss of generality, let us define $Y = \sqrt{X}=\alpha_{l,n_t}\beta_l$. Thus, the corresponding CDF can be given by
\begin{equation}
\begin{aligned}
F_Y(y)&=\Pr(Y\leq y)=\Pr(0\leq X\leq y^2).
\end{aligned}
\end{equation}
Further, the PDF of $Y$ can be evaluated as
\begin{equation}\label{fy1}
\begin{aligned}
f_Y(y)=\frac{d}{dy}F_Y(y)=\frac{d}{dy}\int_0^{y^2}f_X(x)dx.
\end{aligned}
\end{equation}
Furthermore, the (\ref{fy1}) can be calculated as
\begin{equation}
\begin{aligned}
f_Y(y)=2yf_X(y^2)=4yK_0(2y).
\end{aligned}
\end{equation}
Based on this, the mean and variance of $\alpha_{l,n_t}\beta_l$ can be respectively calculated as
\begin{subequations}\label{doumean}
\begin{align}
E[\alpha_{l,n_t}\beta_l]&=\int_0^\infty 4 y^2K_0(2y)dy=\frac{\pi}{4},\\
Var[\alpha_{l,n_t}\beta_l]&=\int_0^\infty 4y^3K_0(2y)dy-\frac{\pi^2}{16}=\frac{16-\pi^2}{16}.
\end{align}
\end{subequations}
By employing the CLT, the proof is completed.

\section{Proof of Theorem 2}
For the sake of representation, we define $x=\xi$.
Since $\xi$ follows the Gaussian distribution, the PDF can be written as
\begin{equation}
f_X(x)=\frac{1}{\sqrt{2\pi\sigma_\xi^2}}\exp\left(-\frac{(x-\mu_\xi)^2}{2\sigma_\xi^2}\right).
\end{equation}
To obtain the distribution of $Y=X^2$, the CDF of event $Y$ can be expressed as
\begin{equation}\label{cdffa}
F_Y(y)\!=\!\Pr(Y\leq y)\!=\!\Pr(X^2\leq y)\!=\!\Pr(-\sqrt{y}\leq X \leq \sqrt{y}).
\end{equation}
Accordingly, we can obtain the PDF via (\ref{cdffa}) as
\begin{equation}\label{cdffa1}
f_Y(y)\!=\!\frac{d}{dy}F_Y(y)\!=\!\frac{1}{\sqrt{2\pi\sigma_\xi^2}}\frac{d}{dy}\!\int_{-\sqrt{y}}^{\sqrt{y}}\!\exp\left(-\frac{(x\!-\!\mu_\xi)^2}{2\sigma_\xi^2}\right)dx.
\end{equation}
By performing the derivative operation on (\ref{cdffa1}), we can get
\begin{equation}\label{cdffa2}
\begin{aligned}
f_Y(y)
=&\frac{1}{2\sqrt{2\pi\sigma_\xi^2}}\frac{1}{\sqrt{y}}\left[\exp\left(-\frac{(\sqrt{y}-\mu_\xi)^2}{2\sigma_\xi^2}\right)
\right.\\&\left.
+\exp\left(-\frac{(-\sqrt{y}-\mu_\xi)^2}{2\sigma_\xi^2}\right)\right].
\end{aligned}
\end{equation}
After some calculations, the (\ref{cdffa2}) can be recast as
\begin{equation}
\begin{aligned}
f_Y(y)
&\!=\!\frac{\exp\left(-\frac{y+\mu_\xi^2}{2\sigma_\xi^2}\right)}{2\sqrt{2\pi\sigma_\xi^2y}}\left[\exp\left(\frac{\mu_\xi\sqrt{y}}{\sigma_\xi^2}\right)\!+\!\exp\left(-\frac{\mu_\xi\sqrt{y}}{\sigma_\xi^2}\right)\right].
\end{aligned}
\end{equation}
At this point, the proof is completed.

\section{Proof of Lemma 1}
Recall that the function $\Phi(x)$, we consider
\begin{equation}
\Phi(-x)=\frac{2}{\sqrt{\pi}}\int_0^{-x}\exp(-v^2)dv.
\end{equation}
By changing the variable of integration, let $u=-v$.
When $v$ goes from $0$ to $-x$, $u$ goes from $0$ to $x$, and $dv = -du$. Therefore,
\begin{equation}
\begin{aligned}
\Phi(-x)&=\frac{2}{\sqrt{\pi}}\int_0^x\exp(-(-u)^2)(-du)\\
&=-\frac{2}{\sqrt{\pi}}\int_0^x\exp(-u^2)du = -\Phi(x).
\end{aligned}
\end{equation}
Since $\Phi(-x)=-\Phi(x)$, $\Phi(x)$ is an odd function.
Substituting $x$ with $\sqrt{\frac{\mu_\xi^2\sin^2\theta}{(2\sin^2\theta+\rho\sigma_\xi^2 |s-\hat s|^2)\sigma_\xi^2}}$, we complete the proof.

\section{Proof of Lemma 2}
For brevity, let us make $x=\theta_{l,\hat n_t},y=\theta_{l,n_t}$. In this case, the corresponding PDF can be formulated as
\begin{equation}
\begin{aligned}
&f_X(x)=\frac{1}{2\pi}, \ \
f_Y(y)=\frac{1}{2\pi}, \ \ \ x,y\in [-\pi,\pi].
\end{aligned}
\end{equation}
Then, we use $z$ to denote the $\phi_l$, the CDF can be given as
\begin{equation}\label{fzxdg1}
\begin{aligned}
F_{Z}(z)\!=\!&\Pr\{Z\!\leq \!z\}
\!=\!\Pr\{X\!-\!Y\leq z\}
\!=\!{\iint}_{D}\!f_{XY}(x,y)dydx,
\end{aligned}
\end{equation}
where $ D: X-Y\leq z $ represents the integration area of $Z$.
Since $X$ and $Y$ are independent of each other, the joint probability density function $f_{XY}(x,y)$ can be written as the product of two marginal probability density functions $f_X(x)$ and $f_Y(y)$, i.e., $f_{XY}(x,y)=f_X(x)f_Y(y)$.
In this way, we can express (\ref{fzxdg1}) as
\begin{equation}\label{fzxdg2}
F_Z(z)=\int_{-\infty}^\infty f_X(x)dx\int_{-\infty}^{x-z}f_Y(y)dy.
\end{equation}
Without loss of generality, let us set $t = x-y$, that is, $dy=-dt$.
As such, the (\ref{fzxdg2}) can derived as
\begin{equation}
F_Z(z)=-\int_{-\infty}^\infty f_X(x)dx\int_{-\infty}^zf_Y(x-t)dt.
\end{equation}
Based on this, the corresponding PDF can be obtained via CDF as
\begin{equation}
f_Z(z)=\frac{d}{dz}F_Z(z)=\int_{-\infty}^\infty f_X(x)f_Y(x-z)dx.
\end{equation}
Since $x\in[-\pi,\pi]$ and $y\in[-\pi,\pi]$, we have $-\pi\leq x-z\leq z$, i.e., $z\in[x-\pi,x+\pi]$.

If $-\pi\leq z+\pi$ and $z-\pi\leq-\pi$, i.e., $z\in[-2\pi,0]$, we have
\begin{equation}
f_Z(z)=\int_{-\pi}^{z+\pi}\frac{1}{4\pi^2}dz=\frac{z+2\pi}{4\pi^2}.
\end{equation}

If $z-\pi\leq \pi$ and $z+\pi\leq\pi$, i.e., $z\in[0,2\pi]$, we have
\begin{equation}
f_Z(z)=\int_{z-\pi}^{\pi}\frac{1}{4\pi^2}dz=\frac{-z+2\pi}{4\pi^2}.
\end{equation}
Herein, the proof is completed.

\section{Proof of Theorem 3}
To obtain the variance of $\gamma_\Im$, we derive the second moment of $\gamma_\Im$ as
\begin{small}
\begin{equation}
\begin{aligned}
E[\gamma_\Im^2]
=& E\left[\sum_{l=1}^{L}\beta_l\left(\alpha_{l,n_t}s_\Im-\alpha_{l,\hat n_t}(\cos\phi_l\hat s_\Im-\sin\phi_l\hat s_\Re)\right) \right]^2\\
\overset{(a)}{=}& \sum_{l=1}^{L}E\left[\beta_l^2\left(\alpha_{l,n_t}s_\Im-\alpha_{l,\hat n_t}(\cos\phi_l\hat s_\Im-\sin\phi_l\hat s_\Re)\right)^2 \right]\\
\overset{(\ref{ere3})}{=}& \sum_{l=1}^{L}E\left[\left(\alpha_{l,n_t}s_\Im-\alpha_{l,\hat n_t}(\cos\phi_l\hat s_\Im-\sin\phi_l\hat s_\Re)\right)^2 \right]\\
{=}& \sum_{l=1}^{L}E\left[(\alpha_{l,n_t}s_\Im)^2+\alpha_{l,\hat n_t}^2(\cos\phi_l\hat s_\Im-\sin\phi_l\hat s_\Re)^2\right.\\&\left.-2\alpha_{l,n_t}s_\Im \alpha_{l,\hat n_t}(\cos\phi_l\hat s_\Im-\sin\phi_l\hat s_\Re) \right]\\
\overset{(b)}{=}& \sum_{l=1}^{L}E\left[(\alpha_{l,n_t}s_\Im)^2+\alpha_{l,\hat n_t}^2(\cos\phi_l\hat s_\Im-\sin\phi_l\hat s_\Re)^2 \right]\\
\overset{(\ref{ere3})}{=}& \sum_{l=1}^{L}E\left[|s_\Im|^2+(\cos\phi_l\hat s_\Im-\sin\phi_l\hat s_\Re)^2 \right]\\
=& \sum_{l=1}^{L}E\left[|s_\Im|^2+\cos^2\phi_l|\hat s_\Im|^2+\sin^2\phi_l|\hat s_\Re|^2\right.\\&\left.-2\cos\phi_l\hat s_\Im\sin\phi_l\hat s_\Re \right]\\
\overset{(b)}{=}& \sum_{l=1}^{L}E\left[|s_\Im|^2+\cos^2\phi_l|\hat s_\Im|^2+\sin^2\phi_l|\hat s_\Re|^2\right]\\
\overset{(c)}{=}& \sum_{l=1}^{L}E\left[|s_\Im|^2+\frac{1+\cos2\phi_l}{2}|\hat s_\Im|^2+\frac{1-\cos2\phi_l}{2}|\hat s_\Re|^2\right]\\
\overset{(\ref{phic1})}{=}& \sum_{l=1}^{L}E\left[|s_\Im|^2+\frac{1}{2}|\hat s_\Im|^2+\frac{1}{2}|\hat s_\Re|^2\right]\\
\overset{(d)}{=}& \left(|s_\Im|^2+\frac{|\hat s|^2}{2}\right)L,
\end{aligned}
\end{equation}
\end{small}%
where $(a)$ denotes the expectation property of the summation of i.i.d. random variables,
$(b)$ represents the use of (\ref{phic1}) and (\ref{phis1}),
$(c)$ indicates the use of the trigonometric function squaring rule, and
$(d)$ stands for the use of CLT.

\end{appendices}

\end{document}